\documentclass[review]{elsarticle}

\usepackage{lineno,hyperref}
\modulolinenumbers[5]

\usepackage{amsmath}
\usepackage{color}
\journal{Journal of \LaTeX\ Templates}

\usepackage{graphicx}% Include figure files
\usepackage{float}% Im­proves the in­ter­face for defin­ing float­ing ob­jects

\begin{document}

\begin{frontmatter}

\title{Anisotropy in interface stress at the {\em bcc}-iron solid-melt interface: molecular dynamics and phase field crystal modelling}
%\tnotetext[mytitlenote]{}

%\author{Sushil Kumar$^1$, Ming-Wei Liu$^2$, Kuo-An Wu$^2$ and M P Gururajan$^1$}
%\address{1. Department of Metallurgical Engineering and Materials Science, Indian Institute of Techb}

\author[IITB]{Sushil Kumar}
\author[NTHU]{Ming-Wei Liu}
\author[NTHU]{Kuo-An Wu}
\author[IITB]{M P Gururajan\corref{CorresAuth}}
\cortext[CorresAuth]{Corresponding author}
\ead{gururajan.mp@gmail.com,guru.mp@iitb.ac.in}

%\author[IITB]{Global Customer Service\corref{mycorrespondingauthor}}

\address[IITB]{Department of Metallurgical Engineering and Materials Science, Indian Institute of Technology Bombay,
Powai, Mumbai 400076 INDIA}
\address[NTHU]{Department of Physics, National Tsing Hua University, Hsinchu 30013 TAIWAN}

\begin{abstract}
The interface stresses at of the solid-melt interface are, in general, anisotropic. The anisotropy in the interfacial
stress can be evaluated using molecular dynamics (MD) and phase field crystal (PFC) models. In this paper, we report 
our results on the evaluation of the anisotropy in interface stress in a {\em bcc} solid with its melt. Specifically,
we study Fe using both MD and PFC models. We show that while both MD and PFC can be used for the evaluation, and the 
PFC and the amplitude equations based on PFC give quantitatively consistent results, the MD and PFC results 
are qualitatively the same but do not match quantitatively. We also find that even though the interfacial free energy 
is only weakly anisotropic in {\em bcc}-Fe, the interfacial stress anisotropy is strong. This strong anisotropy has 
implications for the equilibrium shapes, growth morphologies and other properties at nano-scale in these materials.
\end{abstract}


\end{frontmatter}

%\linenumbers

\section{Introduction}

The properties of the solid-melt interface in crystalline solids is of great fundamental interest and
practical importance in many problems such as solidification and crystal growth~\cite{PorterEasterling}. 
Specifically, the interfacial energy plays a key role in nucleation~\cite{Turnbull1950,AstaEtAl2009} and 
the interfacial energy anisotropy plays a key role in determining (a) the equilibrium shape of nanoparticles 
(see, for example,~\cite{LiuEtAl2001,BackofenVoigt2009}) and (b) the morphology of the growing crystals 
(see, for example,~\cite{Sekerka2005} for a review). 

Several experimental methods (equilibrium shape, grain boundary grooving, nucleation, wetting and dendritic growth)
have been used to measure the interfacial free energy and its anisotropy~\cite{NapolitanoEtAl2002}. However,
given the challenges associated with these experimental studies, theoretical and computational studies
are also very common; see, for example,~\cite{NapolitanoEtAl2002,Davidchack2000,HoytEtAl2001,HoytEtAl2004}. Specifically,
both molecular dynamics (MD) and phase field crystal (PFC) methods have been employed to successfully evaluate
interfacial energies and their aniosotropy; see, for example,~\cite{HoytEtAl2004} for MD studies 
and~\cite{WuKarma2007,Jaatinen2009,Toth2014} for PFC studies.

In a crystalline solid in contact with its melt, one can distinguish interface energy from interface stress; 
the energy associated with the formation of a new interface at constant stress is the interface energy while 
the energy associated with deforming the interface is the interface stress~\cite{LeoSekerka}. Further, the 
properties of a crystalline solid are necessarily anisotropic~\cite{PorterEasterling}, and, hence, like the 
interfacial free energy and interface stress is also known to be anisotropic: see, for 
example~\cite{Herring,Shuttleworth,JohnsonAlexander}. The effect of such interface stress on solid-melt 
equilibrium is well known; see~\cite{LeoSekerka,Eremeyev2016,MomeniLevitas2016} for example.  

Given the difficulties in the experimental determination of interface stress, simulations have been used
in the past decade to determine interface stress. For example, Mishin and Frolov, in a series
of papers, have used MD to determine interface stress in several {\em fcc} 
materials~\cite{Frolov2009,Frolov2010,Frolov2010b,Frolov2010a}. Similarly,
the PFC methodology has been used in to study the interface stress in 2-D hexagonal materials~\cite{LiuWu2017}.

In this paper, we are interested in the evaluation of interfacial stresses and their anisotropy in {\em bcc} solids
using both PFC and MD. Given that PFC can be considered as a time-averaged MD, we want to compare the interfacial stress 
evaluated using these two methods. Further, it is known that the interfacial free energy anisotropy is weak in 
{\em bcc} solids~\cite{WuKarma2007,Jaatinen2009,Wu2006}; hence, we are interested in answering the question, namely, whether 
the  interfacial  stress anisotropy is also weak in such solids. Specifically, we evaluate interface stresses for the 
{\em bcc} iron in contact with its melt; the interface stress is evaluated using molecular dynamics (MD) simulations, the 
phase field crystal (PFC) method and the amplitude equations of the PFC method. 

The rest of the paper is organised as follows: in Section~\ref{Section2}, we describe the MD formulation and simulation
methodologies. The formulation for evaluating interface stresses using MD is well known in the literature; so, we only 
give a brief description for the sake of completion. However, we describe the simulation methodology in detail. In 
Section~\ref{Section3}, we describe the PFC formulation and the amplitude equations 
derived from PFC.  Here again, the PFC formulation is well documented in the literature and hence
we only describe it very briefly for the sake of completion. On the other hand, we extend the amplitude
equation formulation previously for 2D hexagonal lattices \cite {LiuWu2017} to \emph{bcc} solids; hence, we
describe this formulation in greater detail below.  In 
Section~\ref{Section4}, we present the results from our MD and PFC simulations on the interfacial
stress and its anisotropy in {\em bcc} iron in contact with its melt and compare the results from MD with those
obtained from PFC and amplitude expansion. While the PFC and amplitude expansion results agree quantitatively, the
agreement between MD and the PFC/Amplitude equations is qualitative. Finally, we conclude the paper with a brief summary of
salient results from this work.

\section{MD: Formulation and simulation methodology} \label{Section2}

\begin{figure}
\includegraphics[scale=0.3]{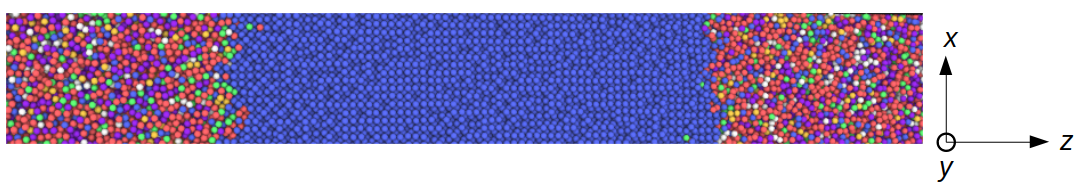}
\caption{A simulation block of solid iron ({\em bcc} -- in blue at the middle of the simulation cell) in contact
with iron melt (on the left and right ends of the simulation cell). The interface normal is along
the z-direction of the simulation cell (which is aligned with the (001) of the crystallographic
direction in this particular case.}\label{InitialCondition}
\end{figure}

Consider the solid-melt interface in a single component system as shown in Fig~\ref{InitialCondition}. 
Our aim is to evaluate the interface stress at this solid-melt interface; specifically,
the interface consists
of solid ({\em bcc}) iron and its melt. In order to evaluate the interface
stress, the solid is non-hydrostatically stressed. The application of non-hydrostatic
stress is known to change the temperature and pressure of the system. In this paper,
we report the interface stresses obtained under isothermal conditions; that is, the solid and 
the melt are maintained at the same temperature both before
and after the application of the non-hydrostatic stress to the solid. There are
other paths, such as the iso-fluid path for example, in which both the pressure
and temperature are maintained to be the same both before and after the application of
the non-hydrostatic stress to the solid. However, here, we restrict
ourselves to the studies carried out using iso-thermal paths. We are also interested
in the anisotropy in the interface stress. Hence, we set up simulations with different
normals for the solid surfaces that are in contact with the melt. 

\subsection{Formulation}

The formulation and methodology used to evaluate surface stress using MD is given in detail in the papers 
of Frolov and Mishin~\cite{Frolov2010, Frolov2010a}. For the sake of completeness in this section
we describe the salient aspects of the formulation relevant for our work. 

Let the square braces $[(\cdot)]$ represent the interface excess quantity in the property $(\cdot)$; then,
the interface excess in the property Z can be represented as a ratio of two determinants~\cite{Frolov2010a} 
as follows:
\begin{eqnarray}
[Z]_{XY}= \frac{\begin{vmatrix}
      [Z] & [X] & [Y] \\ 
      Z^s & X^s & Y^s  \\
      Z^m    & X^m & Y^m  \\ 
      \end{vmatrix}}{\begin{vmatrix}
      X^s      & Y^s \\ 
      X^m      & Y^m \\
      \end{vmatrix}}
      \label{equation-1}
  \end{eqnarray}
where, X and Y are any two extensive properties chosen from energy (U), entropy (S), volume (V) 
and number of atoms (N); the quantities in the first row of the determinant in the numerator are for 
the chosen layer of the interface; for the 
quantities in the other rows of the determinant in the numerator and
the determinant in the denominator, the superscripts $s$ and $m$ represent the homogeneous regions of 
solid and melt phases, respectively. 

For the isothermal path simulations, using the Eq.~\ref{equation-1}, the excess interface stress
$[\tau_{ij}]$ while keeping the volume and number of atoms a constant (that is, choosing
$N$ for $X$ and $V$ for $Y$ in the expression) is as follows:
\begin{eqnarray}
[\tau_{ij}]_{NV}= \frac{\begin{vmatrix}
      [(\sigma_{ij}+\delta_{ij}P)V] & [N] & [V] \\ 
      (\sigma_{ij}^s+\delta_{ij}P)V^s & N^s & V^s  \\
      0    & N^m & V^m  \\ 
      \end{vmatrix}}{\begin{vmatrix}
      N^s      & V^s \\ 
      N^m      & V^m \\
      \end{vmatrix}}
      \label{equation-2}
  \end{eqnarray} 
  
where $\delta_{ij}$ is the Kronecker delta (that is, $\delta_{ij}$ is unity when $i=j$ and zero otherwise),
$\sigma_{ij}$ is the stress and $P$ is the pressure. Expanding the determinants and simplifying, we obtain
\begin{equation} \label{equation-3}
[\tau_{ij}]_{NV} = \frac{1}{A}\left\{[(\sigma_{ij} +   
\delta_{ij}P)V] - (\sigma_{ij}^s + \delta_{ij}P)V^s \left[\frac{[N] V^m - N^m [V]}{N^sV^m - N^mV^s}\right]\right\}
\end{equation} 

Given Eq.~\ref{equation-3}, we can obtain the excess interface stress by carrying out
an isothermal simulation in which the volume ($V$) and the number of atoms $N$ are maintained
a constant; that is , by carrying out a simulation in the $NVT$ ensemble. 
In order to carry out the simulation at constant temperature, first we need to evaluate
the equilibrium coexistence temperature between the solid and melt -- across different
planar interfaces. Note that in the case of stressed solid, the equilibrium temperature
can change; however, since we are using the isothermal path, the temperature is maintained
a constant in our simulations. In addition, in order to stress the solid non-hydrostatically,
we use the stress-control-via-strain-control approach. The simulation methodology consists of 
the following steps. First, we evaluate the equilibrium temperature
of coexistence for the two phases across a planar interface. Second, we determine the elastic constants
of the solid phase; the elastic moduli thus obtained serve as a benchmark for the potential used, to 
validate the phase co-existence (using the virial stress and the compliance tensor), and, to relate
the imposed strains to applied stresses in the solid. Third, we set up a simulation box 
in which the solid and the melt are in co-existence and the solid is stressed by imposing a given strain.
After equilibration, we evaluate the relevant quantities of interest, namely, interface
stresses (using virial stress).

\subsection{Simulation Details}

Our MD simulations are performed using Large-scale 
Atomic/Molecular Massively Parallel Simulator (LAMMPS). One of the crucial inputs to the MD simulation is the potential. In this study,
we have used the EAM potential developed by Mendelev {\em et al.}~\citep{Mendelev2003}; Mendelev {\em et al.} developed this potential to be appropriate
for both the solid and liquid phases of iron; both first principles and experimental data was used in the fitting exercise.

A simulation block of solid-melt system with periodic boundary conditions in all three 
directions is used; in order to align the interface normal to be along the (001), (110) and the (111)
planes of the {\em b.c.c} solid, the following three geometries are used:
\begin{itemize}
\item[(a)] the (100), (010) and (001) directions of the crystal along the x- y- and z-directions of the simulation
box, respectively; 
\item[(b)] the (001), (1$\bar{1}$0) and (110) directions of the crystal along the x-, y- and z-directions of the simulation box, respectively; and,
\item[(c)] (1$\bar{1}$0), (11$\bar{2}$) and (111) directions of the crystal along the x-, y- and z-directions of the simulation box, respectively. 
\end{itemize}
For example, in Fig.~\ref{InitialCondition}, we show the simulation block with the interface aligned along the (001) direction.

In the following sections, we describe the calculation of melting point for these different geometries (using the solid-melt
coexistence method) and the evaluation of elastic moduli of the solid at the melting point (using the direct method).

\subsubsection{Evaluation of equilibrium melting temperature}

The determination of melting temperature is essential for setting up the coexistence of solid-liquid system in the simulations. The melting
temperature is determined in two steps. In the first step, we obtain an approximate melting temperature of the system by carrying out bulk heating 
of  a solid {\em b.c.c} iron block (of 72000 atoms with periodic boundary conditions in all three directions) and identifying the temperature at which 
abrupt change in potential energy is observed as the tentative melting point. In order to refine the melting temperature thus calculated, we use the coexistence method proposed by Morris {\em et al.} which is widely used in the  literature: see for example~\citep{Morris2002,Liu2013,Ramakrishnan2017}. 

\begin{center}
\begin{figure}[h]
\includegraphics[scale=0.3]{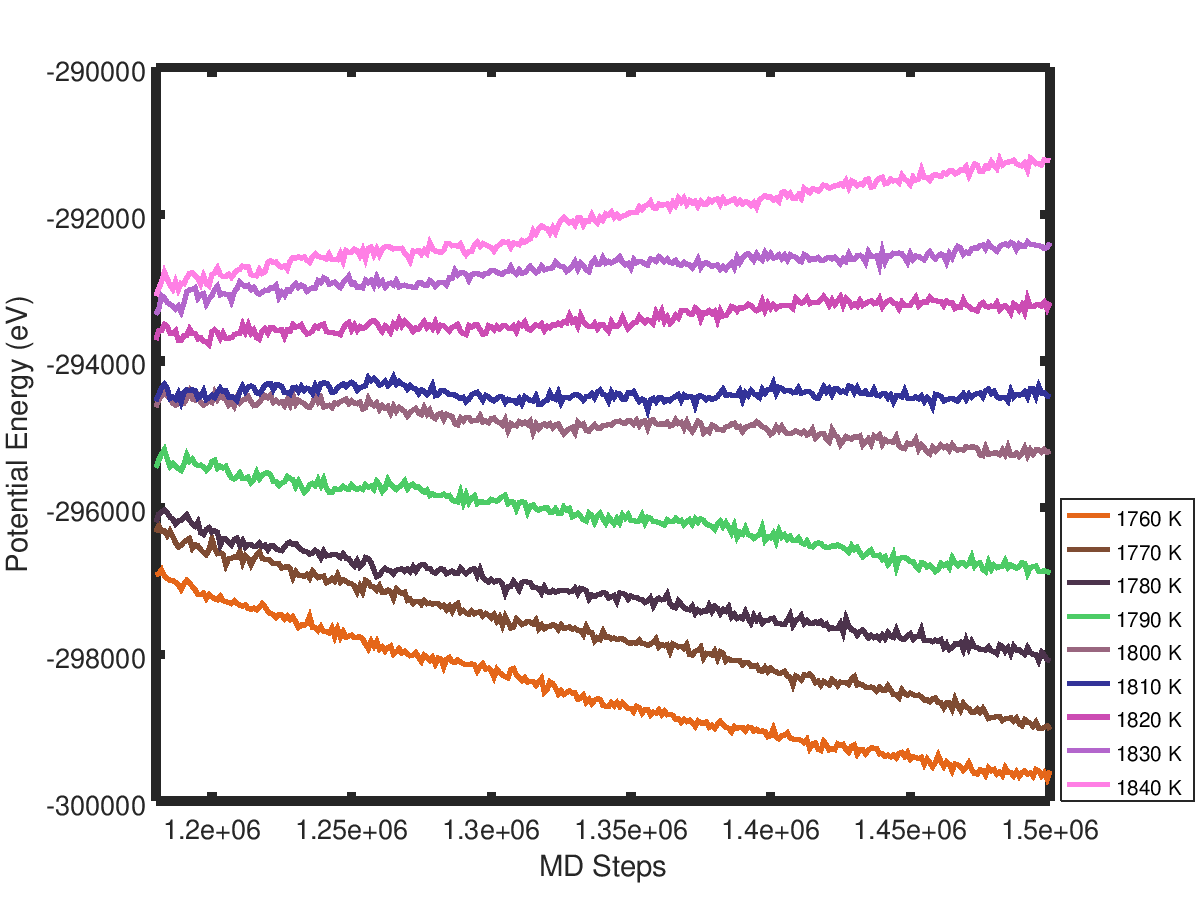}
\includegraphics[scale=0.3]{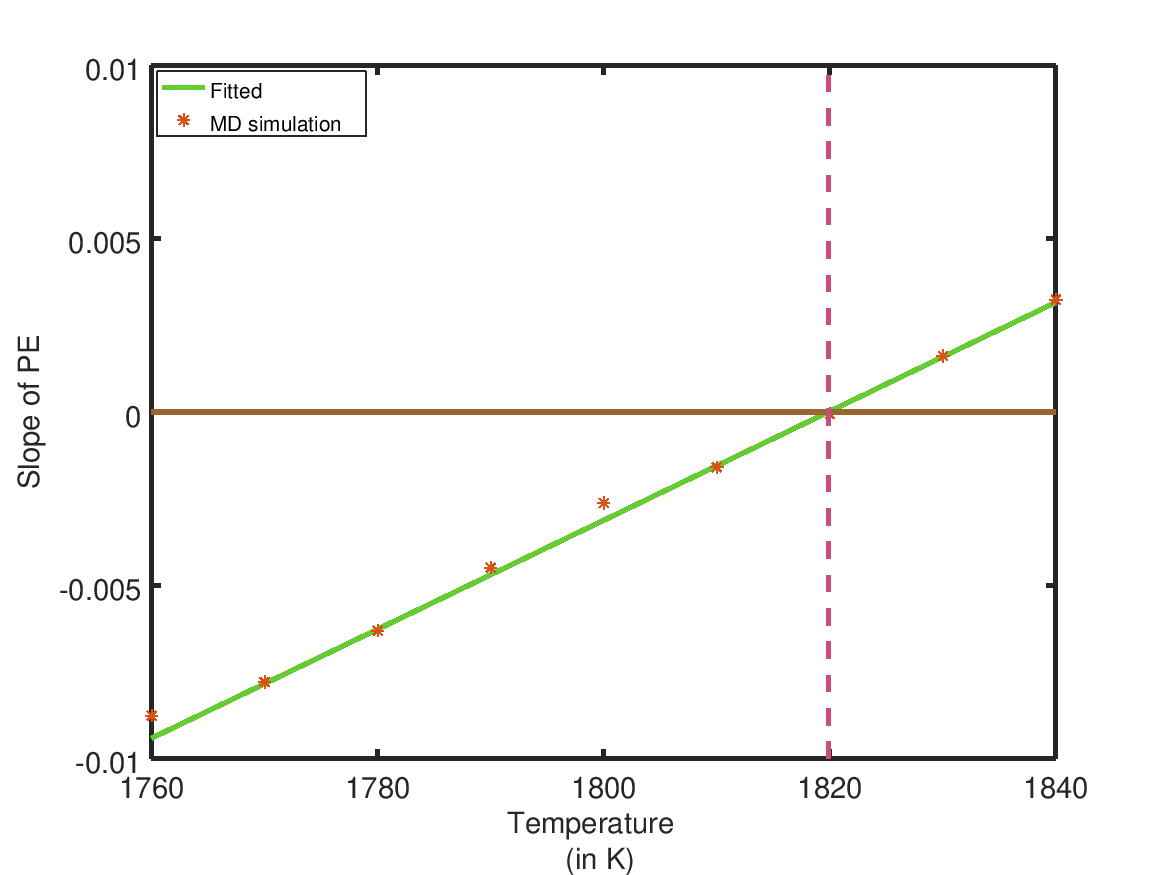}
\caption{Left figure: Potential energy as a function of time at different temperatures obtained using NPT ensemble simulations, and,
Right figure: the slope of the linear portions of the potential energy versus time plots as a function of temperature. The melting
point (dotted vertical line) is the point at which the slope of the potential energy versus time plot is zero.}
\label{potential_energy}
\end{figure}
\end{center}

In the co-existence method, a simulation block consisting of solid and liquid phases is equilibrated at temperatures above and below the estimated melting temperature using the isobaric and isothermal ensemble (NPT) and the potential energy curves are observed during equilibration
at these different temperatures. For example,  for the geometry of (100), (010) and (001) crystallographic directions oriented along the 
x- y- and z-directions, the potential energy curves for different temperatures are as shown in Figure~\ref{potential_energy}. The potential energy 
curve for systems equilibrated below 1820 K decreases, indicating that the system is freezing, while, those equilibrated above 1820 K increase, indicating that the system is melting. At 1820 K, potential energy is a constant indicating the solid-melt coexistence. By plotting the slope of the linear regions of potential energy change (obtained using linear fit) as a function of temperature  (see the second plot in Fig.~\ref{potential_energy}), we identify the point 
where slope of potential energy is zero as the equilibrium melting temperature. Similarly, for other geometries also
the melting temperature is calculated and the results are tabulated in Table.~\ref{MeltingTemperature}. The error bars
are obtained by averaging over three simulations.

%\begin{table}[h!]
%  \begin{center}
%    \caption{Melting temperature for different orientations}
%    \label{tab:table1}
%    \begin{tabular}{l|s|r|t}
%      \toprule % <-- Toprule here
%      \textbf{Orientation} & \textbf{Melting temperature} \\
%      $\alpha$ & $\beta$ & $\gamma$ \\
%      \midrule % <-- Midrule here
%      1 & 1110.1 & a\\
%      2 & 10.1 & b\\
%      3 & 23.113231 & c\\
%      \bottomrule % <-- Bottomrule here
%    \end{tabular}
%  \end{center}
%\end{table}
\begin{table}[h!]
  \begin{center}
    \begin{tabular}{l|c|c|c|c} % <-- Alignments: 1st column left, 2nd middle and 3rd right, with vertical lines in between
    \hline
Interface plane  & X & Y & Z & Equilibrium melting temperature \\
     orientation & &  & & K \\
      \hline
       (001) & [100] & [010] & [001] &  1820 $\pm$ 1 \\
       (110) & [001] & [1$\overline{1}$0] & [110] & 1817 $\pm$ 3\\
       (111) & [1$\overline{1}$0] & [11$\overline{2}$] & [111] & 1820 $\pm$ 2\\
      \hline
    \end{tabular}
     \caption{Equilibrium melting temperature for different geometries across a planar solid-melt interface obtained
     using the coexistence method.}\label{MeltingTemperature}
  \end{center} 
\end{table}

\subsubsection{Calculation of elastic constants}

The elastic constant is computed using the direct method and simulations in an NVT ensemble -- as described
in the LAMMPS manual~\cite{LAMMPSManual}. We use a strain of 0.2\% .
The elastic constants at 0 K, room temperature (300 K) and at the melting temperature ($T_m = 1820$ K) obtained
using our computations are given in Table~\ref{table :1}. These compare well with the constants
reported for 0 K using first principle calculations (as reported by Mendelev {\em  et al.}~\cite{Mendelev2003}, 
the experimentally reported constants at 300 K~\cite{Rayne1961} and the near melting point constants 
reported~\cite{Etesami2018} (albeit using a different potential and different method of evaluation).
\begin{table}
\begin{tabular}{l c c c c}
\hline
&Elastic constants & $ C_{11}$ & $C_{12}$ & $C_{44}$\\[0.5ex]
  & & (GPa) & (GPa) & (GPa)\\
\hline
Current& 0 K & 243&131 &115 \\
study & 300 K & $230 \pm 2$ & $134 \pm 2$ & $105 \pm 5$\\
& 1820 K & 132 & 103 & 73\\
\hline
Reported& 0 K (DFT) &243.7 & 145.1 & 115.9\\
Values& 300  K (Expt) & 234 &131 & 115\\
& 1812 K (MEAM potential / MD simulations) &141 &129.2 &68.1 \\
\hline
\end{tabular}
\caption{Elastic constants of Fe from the current simulations compared with reported values in the literature.}
\label{table :1}
\end{table}

\subsubsection{Interface stresses in solid-melt interface at constant temperature}

Once the solid-melt coexistence is achieved, the simulation block is subjected to biaxial deformation by applying equal 
magnitude of stress in x and y directions to generate non-hydrostatic stresses in the solid by scaling the dimensions 
of the simulation block. The simulation block is subjected to both tension and compression biaxially. The deformation 
destroys the phase equilibrium and hence the coexisting phases are again re-equilibrated in the canonical ensemble (NVT) 
for 2 million time steps after deformation; this is followed by production run of 4 million time steps and the results 
of the simulation are saved after every 10000 time steps. In our simulations each time step corresponds to 1 femtosecond. 
The interface stresses are obtained using Eq.~\eqref{equation-3} by analysing the saved results and generating the 
required quantities (such as volume, atomic stress etc) from the simulation results.\par

\subsubsection{Validation of phase coexistence}

\begin{figure}[h!]
\begin{center}
\includegraphics[scale=0.8]{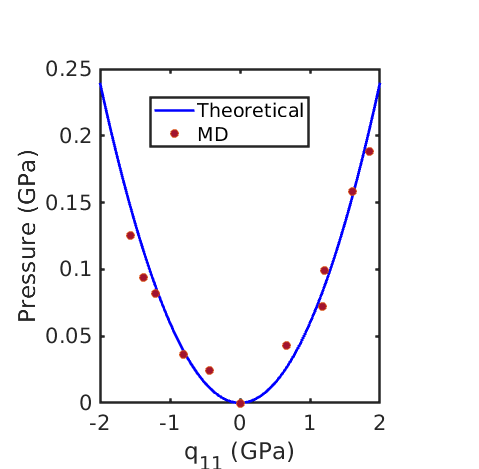}
\end{center}
\caption{Equilibrium pressure of the liquid as a function of the non-hydrostatic stress component in the solid (which is
subjected to biaxial stress).}
\label{ExcessPressure}
\end{figure}

When the solid is deformed non-hydrostatically, the pressure in the melt changes and the change in the pressure is given
by the following analytical expression:
\begin{equation}
P - P_H = \frac{V^{s}}{2 (V^{m}-V^{s})}[S_{1111}q_{11}^2 + 2S_{1122}q_{11}q_{22} + S_{2222}q_{22}^2]
\label{theoreticalpressure}
\end{equation}
where $P_H$ is hydrostatic pressure in the absence of any applied stress, $S_{ijkl}$ is the compliance tensor, 
$V^s$ and $V^m$ are the volumes of the homogeneous solid and melt respectively, and $q_{ij}$ are the components of 
the virial stress (that is, $\sigma_{ij} + p \delta_{ij}$) -- see 
Eq. 37 of ~\citep{Frolov2010b} (Note that Frolov and Mishin's version also contains a $T_H$ term in the numerator 
on the RHS which, we believe, is a typographical error). In Fig.~\ref{ExcessPressure}, we compare our numerical 
results for the $P-P_H$ obtained using MD simulations with that given by the analytical expression 
above and the two match very well indicating that our simulations are in agreement with the analytical prediction. 
Note that the given figure is for one of the geometries, namely, x-, y- and z-directions of the simulation
cell are aligned with the (100), (010), and (001) crystallographic directions respectively; we find similar agreement 
for the other two geometries also.

\subsubsection{Identification of the interface position}

In order to calculate the interface stress components, we need to evaluate the extensive quantities for homogeneous 
bulk solid and melt regions as well as the interface region. Hence, we need to evaluate the position of the interface and 
its width; the homogeneous bulk solid and melt regions are identified with respect to the instantaneous position of the 
interface. Hence, the determination of the position of the interface and its width are crucial. 

Due to thermal fluctuations, the position of the interface keeps changing with time during the MD run. In order to locate 
the instantaneous interface position, a structural order parameter ($\phi_k$), is calculated using the given equation:
\begin{equation}
\phi_k = \frac{1}{14} \sum_i |r_i - r_{bcc}|^2
\label{orderparameter}
\end{equation}
where r$_i$ denotes the distances of the first and second nearest neighbours 
in the simulated structure and r$_{bcc}$ refers to the first and second nearest neighbour distance for the ideal {\em bcc}
 crystal. This definition of order parameter 
 is known to result in  $\phi_k \approx 0$ for an atom in the {\em bcc} solid 
and non-zero for atoms in the melt regions: see for example~\citep{Ramakrishnan2017,HoytEtAl2001}. Thus, the $\phi_k$,
defined for every atom $k$ helps us distinguish the solid and liquid regions.
  
\begin{figure}[htbp]
\begin{center}
\includegraphics[scale=0.6]{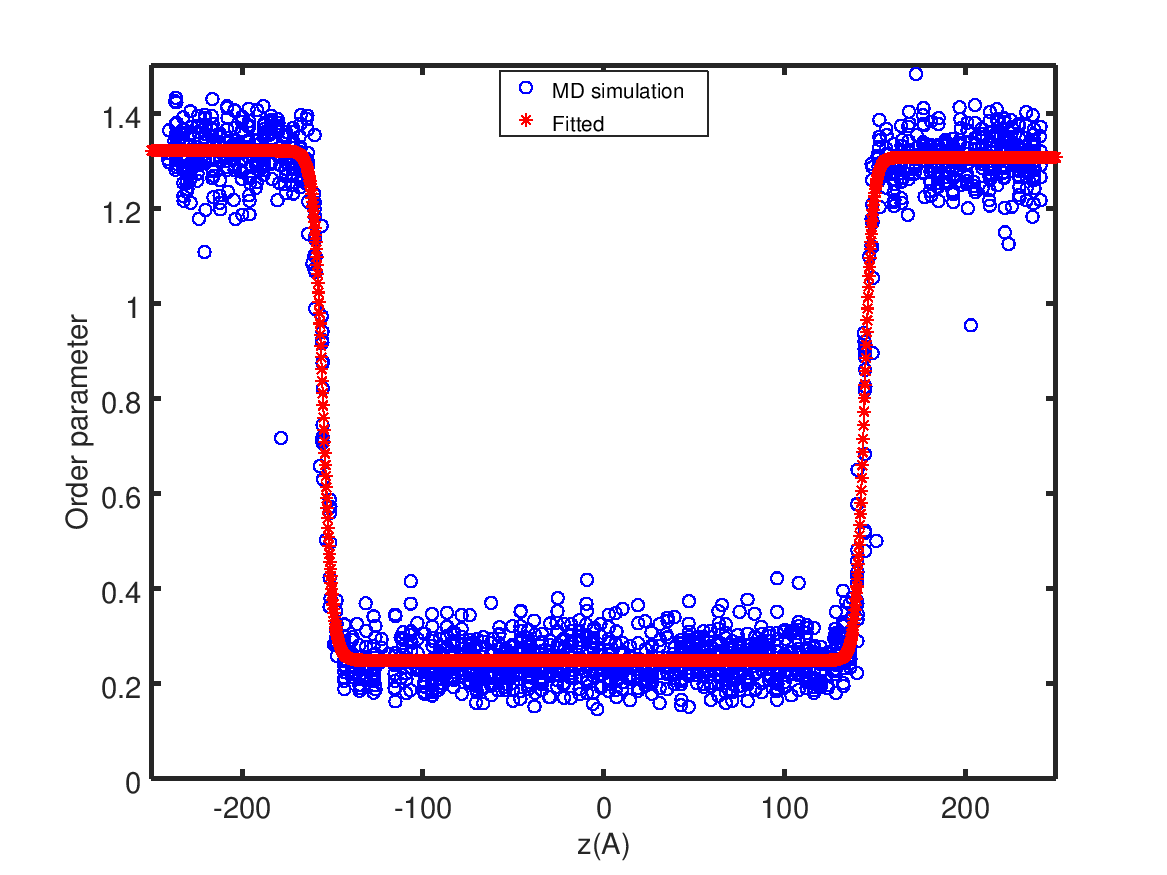}
\end{center}
\caption{Calculated order parameters to determine interface position with respect to a coordinate perpendicular to 
the solid-melt interface.}
\label{order_para}
\end{figure}

The order parameter $(\phi_k)$ is further smoothed using the following equation to remove the fluctuation in a calculated values: 
\begin{equation}
\phi(z_p) = \frac{\sum_k w(z_p-z_k)\phi_k}{\sum_k w(z_p-z_k)}
\label{weightingfunction}
\end{equation}
where weighting function $w$ is defined as $w(z_p) = \left[1-(\frac{z_p}{d})^2 \right]^2$ for $|z_p| < d$ (cut-off distance), 
else $w(z_p) = 0$ and $z_k$ is atom coordinates of atom $k$. In the z-direction the simulation block is discretised into bins 
and order parameters are averaged over all atoms in the bin; we use a cut-off distance of $7\AA$ in order to avoid the smearing 
out of the information~\citep{Davidchack2006}. The position of the interface at any instant is determined by fitting a function 
$f(z) = c_1 + c_2 \tanh\left( \frac{z-c_3}{c_4} \right)$  to the order parameter data ($\phi(z_p)$), where z is the coordinates 
of the atoms along the direction perpendicular to the solid-melt interface and the fitting parameter c$_3$ denotes the interface
 position~\citep{Ramakrishnan2017}.  The plot of the order parameters along with the fit of the hyperbolic tangent function is 
 shown in Figure~\ref{order_para}. 

In order to make sure that the integration for the interface properties is done over the entire interface, we have to choose
an interface width (`d') which extends from bulk melt to bulk solid or vice versa. In this study, we have used a $d$ of 
20$\AA$.

\section{Phase field crystal model: Formulation and simulation methodology} \label{Section3}

The PFC model has been widely used to investigate elastic and plastic processes in materials as well as solid-solid and solid-liquid interfacial properties \cite{Elder2002,Elder2004,Pisutha-Arnond2013,Heinonen2014,Wu2009,Yu2011,Choudhary2014,Podmaniczky2017, Backofen2010,Toth2010,Tang2014}. It has its roots in the density functional theory of freezing as discussed in detail in Ref.~\cite{Elder2007}. Variants of the PFC model have been proposed to generalise the model to describe various crystal symmetries \cite{Wu2010,Greenwood2010,Greenwood2011}. Here, we restrict our attention to the simplest formulation of PFC since we focus on the relation of the surface stress and the underlying crystal symmetry, although the following derivation could be easily generalised for other PFC models. Recently, Liu and Wu have investigated the surface stresses  of the 2D hexagonal lattice \cite{LiuWu2017}. Based on the formalism and calculation in their work, we investigate the interfacial stress for {\em bcc}-liquid interface.

\subsubsection{Free energy functional}

In Ref.~\cite{Elder2002,Elder2004}, the dimensionless form of the simplest PFC free energy functional is
\begin{align}
F = \int d\vec{r} ~\left\{\frac{\psi}{2}\left[-\epsilon +(1+\nabla^2)^2 \right]\psi +\frac{\psi^4}{4}\right\}
\label{Eq_PFC:FreeEnergy_PFC}
\end{align}
in which $\psi$ serves as the dimensionless density field relative to a reference state, and $\epsilon$ is a control parameter that determines the interfacial properties and elasticity of materials which can be set once the liquid structure factor of the material at the melting point is known \cite{WuKarma2007, Elder2004}. The chemical potential $\mu$ is defined as the functional derivative of  $F$, namely
\begin{align}
\mu = \frac{\delta F}{\delta \psi}
= \left[-\epsilon +(1+\nabla^2)^2\right]\psi +\psi^3.
\label{Eq_PFC:ChemicalPotential_PFC}
\end{align}
The dynamics of the system simply follows the following conserved relaxation rule, ${\partial \psi/ \partial t} = \nabla \cdot \nabla \mu$.
According to thermodynamics, the chemical potential $\mu$ must be a uniform constant over the system at equilibrium.

\subsubsection{Solid-liquid coexistence}

PFC model would naturally form the liquid and solid coexistence by setting appropriate  value of parameter $\epsilon$ and the mean density $\bar{\psi}$ \cite{WuKarma2007}. The liquid phase is a homogeneous state, $\psi(\vec{r}) = \bar{\psi}_l$. The corresponding free energy density is obtained straightforwardly from Eq.~\eqref{Eq_PFC:FreeEnergy_PFC},
\begin{align}
f_l = (-\epsilon +1)\frac{\bar{\psi}_l^2}{2} +\frac{\bar{\psi}_l^4}{4}.
\label{Eq_PFC:LiquidPhase_PFC}
\end{align}
In comparison, the solid phase could be expressed approximately by a mean density $\bar{\psi}_s$ and a group of density waves that reflects the symmetry of the solid,
\begin{align}
\psi_s(\vec{r}) \approx \bar{\psi}_s + \sum_j A_j e^{i\vec{K}_j\cdot\vec{r}} +\sum_j A_j^* e^{-i\vec{K}_j\cdot\vec{r}},
\label{Eq_PFC:SolidPhase_PFC}
\end{align}
where $A_j$ and $A_j^*$ are the amplitude and its complex conjugate, respectively, of each density wave, and $\vec{K}_j$ is the reciprocal lattice vector (RLV). Based on the one-mode approximation, the solid state at coexistence can be well represented by employing only the set of principal RLVs. The reciprocal lattice to a {\em bcc} structure is a {\em fcc} structure. Therefore, the principal RLVs for a {\em bcc} structure is a group of 12 vectors, namely,
\begin{align}
\vec{K}_1 &= \frac{K_0}{\sqrt{2}}\left(\hat{x}+\hat{y}\right), 
\qquad \vec{K}_2 = \frac{K_0}{\sqrt{2}}\left(\hat{x}-\hat{y}\right),
\qquad \vec{K}_3 = \frac{K_0}{\sqrt{2}}\left(\hat{y}+\hat{z}\right),\\
\vec{K}_4 &= \frac{K_0}{\sqrt{2}}\left(\hat{y}-\hat{z}\right),
\qquad \vec{K}_5 = \frac{K_0}{\sqrt{2}}\left(\hat{x}+\hat{z}\right),
\qquad \vec{K}_6 = \frac{K_0}{\sqrt{2}}\left(\hat{x}-\hat{z}\right),
\label{Eq_PFC:The1stRLV_PFC}
\end{align}
and their counterparts $\vec{K}_{\bar{j}} = -\vec{K}_j$. Since the crystal symmetry of the {\em bcc} lattice ensures that the all density waves have the same amplitude $A_s$, the free energy density of the solid is readily derived from Eq.~\eqref{Eq_PFC:FreeEnergy_PFC},
\begin{align}
f_s &= (-\epsilon +1)\frac{\bar{\psi}_s^2}{2} +\frac{\bar{\psi}_s^4}{4}
+6[-\epsilon+(1-K_0^2)^2]A_s^2 +18\bar{\psi}_s^2A_s^2 +48\bar{\psi}_sA_s^3 +135A_s^4,
\label{Eq_PFC:SolidFreeEnergy_PFC}
\end{align}
where $K_0 = 1$ so that free energy is minimised with respect to $K_0$. By minimising the free energy density with respect to $A_s$, we obtain the amplitude of density waves of the solid phase, 
\begin{align}
A_s = -\frac{2}{15}\bar{\psi}_s +\frac{1}{15}\sqrt{5\epsilon -11\bar{\psi}_s^2}
\end{align}
which is coupled to $\epsilon$ and $\bar{\psi}_s$.

The coexistence densities for solid-liquid systems with planar interfaces are determined by requiring both phases have the same chemical potential and pressure. Detailed calculations are discussed in Ref. \cite{WuKarma2007}. For small values of $\epsilon$, the coexistence densities can be expressed as a series expansion of $\epsilon^{1/2}$,
\begin{align}
\bar{\psi}_s &\approx \epsilon^{1/2}\psi_{s0} +\epsilon^1\psi_{s1} +\epsilon^{3/2}\psi_{s2} +\mathcal{O}(\epsilon^2)\\
\bar{\psi}_l &\approx \epsilon^{1/2}\psi_{l0} +\epsilon^1\psi_{l1} +\epsilon^{3/2}\psi_{l2} +\mathcal{O}(\epsilon^2).
\end{align}
For {\em bcc} lattices, the coefficients are computed using the common tangent construction which leads to $\psi_{s0} = \psi_{l0} = \psi_c \equiv -\sqrt{45/103}$ and $\psi_{s1}=\psi_{l1} = 0$. Notably, the miscibility gap $\bar{\psi}_s-\bar{\psi}_l$ is proportional to $\epsilon^{3/2}$.

\subsubsection{Interfacial energy}

The interfacial energy is the excess free energy of forming an interface \cite{Frolov2009}, and in the PFC model the interfacial energy of a planer interface can be evalutated by the following expression shown in Ref. \cite{WuKarma2007},
\begin{align}
\label{Eq_PFC:intf_en}
\gamma = \Sigma^{-1} \int d\vec{r}\left[f 
-\left(\frac{\psi-\bar{\psi}_l}{\bar{\psi}_s-\bar{\psi}_l}f_s+\frac{\psi-\bar{\psi}_s}{\bar{\psi}_l-\bar{\psi}_s}f_l\right)\right],
\end{align}
where $\Sigma$ is the area of the planar interface and $f$ represents the free energy density, which is the integrand in Eq.~\eqref{Eq_PFC:FreeEnergy_PFC}.

\subsection{Amplitude Equations for PFC model}
In addition to the PFC model, the amplitude equations (AEs) are often employed to show how the density waves break its symmetry at the interface which eventually leads to anisotropic properties of the interface.
The amplitude equations are obtained using the multi-scale expansion of the PFC model, assuming that the amplitude profile varies much more slowly than the underlying periodicity of the crystal. As shown in Ref.~\cite{WuKarma2007}, the amplitude of density waves depends on a slow spatial variable $\vec{R} = \epsilon^{1/2}\vec{r}$, and thus, we can express the density field as
\begin{align}
\psi(\vec{R},\vec{r}) \approx \epsilon^{1/2}\left[\psi_c + \epsilon n(\vec{R}) 
+\sum_{j=1}^6 A_j(\vec{R}) e^{i\vec{K}_j\cdot\vec{r}} +\sum_{j=1}^6 A_j^*(\vec{R}) e^{-i\vec{K}_j\cdot\vec{r}}\right],
\label{Eq_PFC:multiscale_expansion}
\end{align}
where $n(\vec{R})$ is the average density over a unit cell. As a consequence, the effective free energy functional with respect to amplitudes and average density becomes
\begin{align}
\label{Eq_PFC:free_en_ae}
F &\cong \epsilon\int d\vec{R} ~ \sum_{j=1}^6 |\hat{L}_jA_j|^2 +\bar{f}_{local} +\mathcal{O}(\epsilon^2),\\
\hat{L}_j &\equiv 2i\vec{K}_j\cdot\nabla_R+\epsilon^{1/2}\nabla_R^2,\\
\bar{f}_{local} &\equiv (-1 +\epsilon^{-1})\frac{\psi_c^2}{2} +\psi_c(-\epsilon+1) n +\epsilon\frac{n^2}{2}+\frac{\psi_c^4}{4}+\epsilon\psi_c^3n\nonumber\\
&+(3\psi_c^2 +6\epsilon\psi_cn -1)\sum_{j=1}^3|A_j|^2 +6(\psi_c+\epsilon n)(A_1A_2A_3+A_1^*A_2^*A_3^*)\nonumber\\
&+\frac{3}{2}\sum_{j=1}^3|A_j|^4+6\sum_{j=1}^3\sum_{k>j}^3|A_j|^2|A_k|^2,
\end{align}
in which $\nabla_R$ denotes the gradient with respect to $\vec{R}$. The second term in Eq.~\eqref{Eq_PFC:free_en_ae} is a non-linear function of free energy density which simply depends on local density and amplitudes. Furthermore, the amplitudes $A_j$ and average density $n$ follow the non-conserved and conserved relaxational dynamics, respectively,
\begin{eqnarray}
\frac{\partial A_j}{\partial t} &=& -\frac{\delta F}{\delta A_j^*} 
= -\hat{L}_j^2A_j -\frac{\partial \bar{f}_{local}}{\partial A_j^*}
\label{Eq_PFC:eom_amp_ae}\\ 
\frac{\partial n}{\partial t} &=& \nabla^2\frac{\delta F}{\delta n}
= \epsilon\nabla_R^2\frac{\partial \bar{f}_{local}}{\partial n}
\label{Eq_PFC:eom_dens_ae}.
\end{eqnarray} 
Due to the dependence of $\hat{L}_j$ on $\vec{K}\cdot\nabla_R$, the profile of amplitudes would vary differently across the interface depending on surface orientation which leads to anisotropic interfacial properties, as shown by Wu and Karma \cite{WuKarma2007}. 

In bulk phases, $n$ and $A_j$ are uniform and the free energy densities, $\bar{f_s}$ and $\bar{f}_l$, of solid and liquid, respectively, are expressed by $\bar{f}_{local}(n,A_j)$ with different average density and amplitudes,
\begin{align}
\bar{f}_s = \bar{f}_{local}(n_s, \bar{A}), \qquad
\bar{f}_l = \bar{f}_{local}(n_l, 0)
\label{Eq_PFC:bulk_FE_ae}
\end{align}
where $n_s$ and $n_l$ are the average density of solid and liquid, respectively, and $\bar{A}$ is the common amplitude among the density waves, which corresponds to $A_s$ in PFC, $A_s\cong \epsilon^{1/2}\bar{A}$. Similar to the case of the original PFC model, the values of $n_s$, $n_l$ and $\bar{A}$ at solid-liquid coexistence are solved by common tangent construction to $n_s$ and $n_l$ and free energy minimisation with respect to $\bar{A}$.

\subsubsection{Deformation}

Before deriving interfacial stress for the PFC model, it is necessary to identify the measure of the deformation such as displacement vector $\vec{u}$ in the PFC model, which has been discussed in Refs. \cite{Chan2009,Elder2010,Pisutha-Arnond2013,Heinonen2014}. The deformation could be regarded as a transformation of lattice basis defined as
\begin{align}
\label{trans_basis}
e^{i\vec{K}\cdot\vec{r}} \rightarrow e^{i\vec{K}\cdot\vec{r}\,'} 
=e^{i\vec{K}\cdot(\vec{r}-\vec{u})} 
\end{align} 
which corresponds to shifting a lattice point $\vec{r}\,'$ to a new position $\vec{r}$ with a displacement $\vec{u}$. Then, the phase of $A_j$'s is associated with the local displacement field $\vec{u}$.
This description works well within bulk solid, where the crystal symmetry is preserved; however, it is not sufficient for the interface region where atoms subjected to asymmetric interatomic potential. Therefore, an additional phase, $\varphi$, in the amplitudes is required to completely capture lattice deformations due to asymmetric forces experienced by atoms at the interface. The general form of the amplitude can be expressed as   
\begin{align}
A_j = a_je^{-i\vec{K}_j\cdot\vec{u}}e^{i\varphi}
\label{Eq_PFC:amp_phase}
\end{align}
where $a_j\equiv|A_j|$ is the magnitude of amplitudes and  $\varphi$ is the mean phase of the three amplitudes, $A_1A_2A_3 = |A_1A_2A_3|e^{3i\varphi}$, which is an additional degree of freedom apart from the displacement vector. For a stress-free solid-liquid system at equilibrium, $\vec{u}$ is homogeneous and $\varphi$ vanishes in bulk solid but both of them vary across the interface and their variations are in the order of $\epsilon$, $\nabla\vec{u}\sim\nabla\varphi\sim\epsilon$.

With Eq.~\eqref{Eq_PFC:amp_phase}, one can readily quantify the excess free energy contribution due to strains by applying $\hat{L}_j$ to the amplitudes,
\begin{align}
&\hat{L}_jA_j = \hat{L}_j a_j e^{-i\vec{K}_j\cdot\vec{u}+i\varphi}
\cong e^{-i\vec{K}_j\cdot\vec{u}+i\varphi}\hat{L}_j'a_j +\mathcal{O}(\epsilon^2)\label{Eq_PFC:Lj_strn}\\
&\hat{L}_j' \equiv \hat{L}_j
+2\epsilon^{-1/2}K_{j\alpha}K_{j\beta}\varepsilon_{\alpha\beta}
-2\epsilon^{-1/2}K_{j\alpha}\varphi_{,\alpha}\nonumber\\
&\qquad\qquad-2iK_{j\alpha}u_{\alpha,\beta}\bar{\partial}_\beta -iK_{j\alpha}(\bar{\partial}_\beta u_{\alpha,\beta})
+2i\varphi_{,\alpha}\bar{\partial}_{\alpha} +i(\bar{\partial}_{\alpha}\varphi_{,\alpha})
\end{align}
where $K_{j\alpha}$ represents the $\alpha$ component of the vector $\vec{K}_j$, $\varepsilon$ serves as the strain field, $\varepsilon_{\alpha\beta} \equiv \frac{1}{2}(u_{\alpha,\beta}+u_{\beta,\alpha})$, $u_{\alpha,\beta} \equiv \frac{\partial u_\alpha}{\partial r_\beta}$, $\varphi_{,\beta} \equiv \frac{\partial \varphi}{\partial r_\beta}$ and $\bar{\partial}_{\alpha}\equiv \frac{\partial}{\partial R_{\alpha}}$. For small value of $\epsilon$, we neglect higher order terms of $\epsilon$ as well as higher order terms of $\varepsilon$ in Eq.~\eqref{Eq_PFC:Lj_strn} in the limit of the diffuse interface and linear elasticity.

\subsubsection{Interfacial energy and interfacial stress}

For amplitude equations, the 2D solid-liquid system with a planar interface can be simplified further to an 1D problem due to homogeneity in the direction parallel to the interface. Similar expression of the interfacial energy is derived for AE,
\begin{eqnarray}
\label{Eq_PFC:intf_en_ae}
\gamma &=& \epsilon^{3/2}\int dZ \left[\sum_{j=1}^3 \Big|\hat{L}_jA_j\Big|^2
+\bar{f}_{local} -\left(\frac{n-n_l}{n_s-n_l}\bar{f}_s+ \frac{n_s-n}{n_s-n_l}\bar{f}_l\right)\right]
\end{eqnarray}
in which $\hat{L}_j$ is reduced to one dimensional form, $\hat{L}_j = (\vec{K}_j\cdot\hat{n})\partial_Z +\epsilon^{1/2}\partial_Z^2$, and $Z$ is along the direction of the interface normal $\hat{n}$.

The dependence of the excess interfacial energy on the strains can be shown by substituting Eq.~\eqref{Eq_PFC:Lj_strn} into Eq.~\eqref{Eq_PFC:intf_en_ae},
\begin{equation}
\gamma = \epsilon^{3/2}\int dZ \left\lbrace\sum_{j=1}^3 \Big|\hat{L}_j'a_j\Big|^2 
+\bar{f}_{local} -\left(\frac{n-n_l}{n_s-n_l}\bar{f}_s+ \frac{n_s-n}{n_s-n_l}\bar{f}_l\right)\right\rbrace
\label{Eq_PFC:intf_en_strn_ae}.
\end{equation}
The expression of the interfacial stress is readily obtained using the Shuttleworth equation, $\tau_{tt} = \gamma +\frac{\partial\gamma}{\partial\varepsilon_{tt}}$,
\begin{align}
\tau_{tt}\cong \gamma +\epsilon^{3/2}\int dZ 
\left[\sum_{j=1}^3 4K_{jt}^2a_j\left(\partial_Z^2+2\epsilon^{-1}K_{j\alpha}K_{j\beta}\varepsilon_{\alpha\beta}-2\epsilon^{-1}K_{j\alpha}\varphi_{,\alpha}\right)a_j\right]
\label{Eq_PFC:excess_surf_strs_ae}.
\end{align}
It is clear that the interfacial stress is closely related to the change in the RLVs across the interface as shown above.

\section{Results and discussion} \label{Section4}

We first describe the results from MD simulations, followed by those obtained from PFC and amplitude equations. This section ends
with a qualitative and quantitative comparison of the results from MD and from PFC and amplitude equations. 

\subsection{Results from MD}

The stress components obtained from the MD simulations are the the virial stress (per atom); we use these to evaluate the interfacial 
stress. However, the stress components obtained from the MD simulations are noisy; hence,  we need to smooth the stress profiles 
obtained from the MD simulations before calculating the interfacial stress. We use finite impulse response (FIR) filter method to
carry out the smoothing. 

\begin{figure}[h!]
\begin{center}
\includegraphics[scale=0.3]{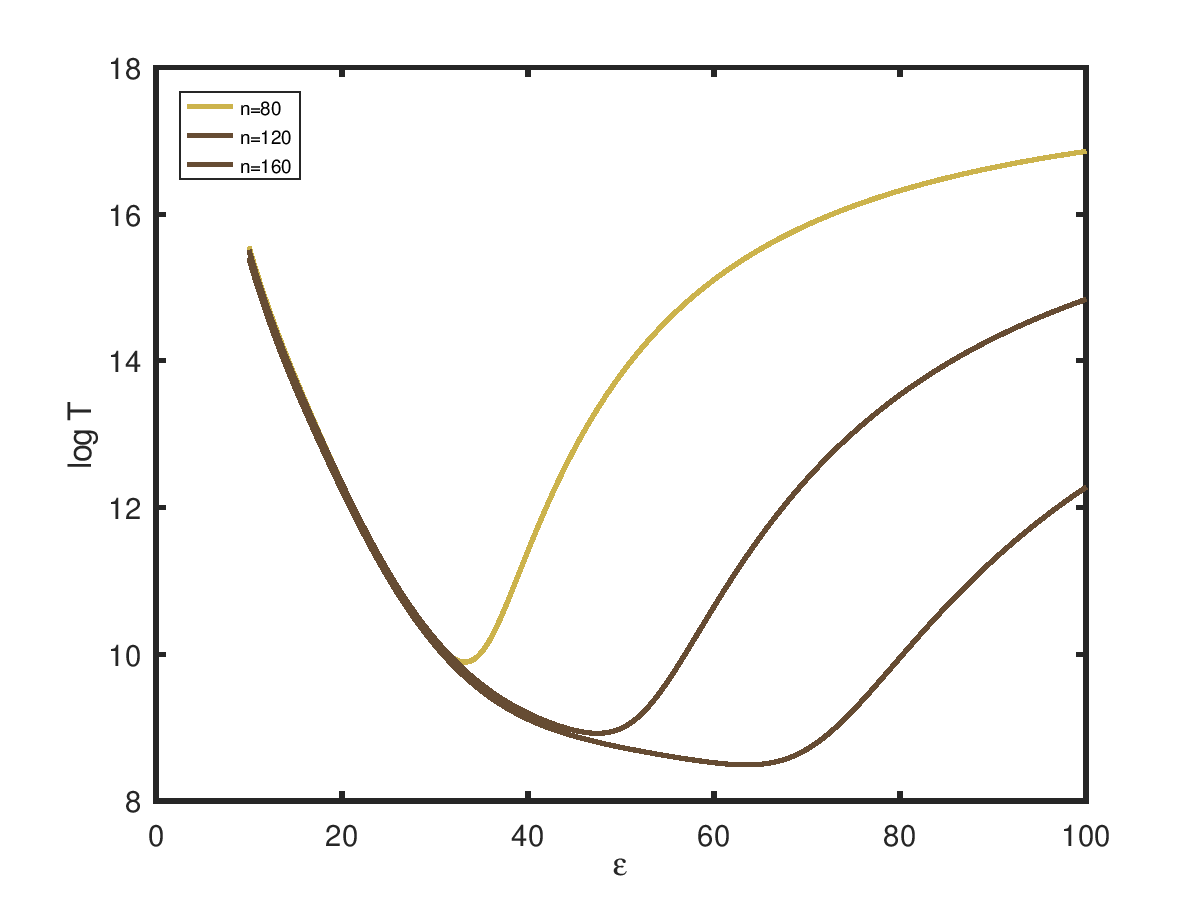}
\end{center}
\caption{$log(T)$ {\em vs} $\varepsilon$ plot to determine the constants of FIR filter.}
\label{FIR}
\end{figure}

We briefly describe the FIR filter method here. This method is based on~\cite{Davidchack1998,yunfei2012} and
interested reader may refer to the same for the details.
We generate the fine scale stress profile by dividing the data in to discrete 
bins along $z$-direction and averaging the data in the $x-y$ plane for every z-value.
 Let $\sigma_{i}$ be the fine scale stress at the point $i$ along the z-axis.
The smoothed stress $\overline{\sigma_i}$ is given by the weighted average
of $2n+1$ data points ($n$ on either side of the position $i$) as follows: 
\begin{equation}
\overline{\sigma_i} = \sum_{k=-n}^{k=n} w_k \sigma_{i+k}
~\label{fir}
\end{equation}
where $w_k$ are the filter coefficients. The $w_k$ is of Gaussian form $w_k = Ae^{-(k/\epsilon)^2}$ 
where $A$ is normalization constant and is calculated from the condition $\sum w_k = 1$. The values of 
$w_k$ are determined by minimising, $T = \sum_n (\delta^2 \overline{\sigma_i})^2$ where $\delta^2$ 
is the second central difference formula, that is, 
$\delta^2 \overline{\sigma_i} = {\overline \sigma_{i+1}} + {\overline \sigma_{i-1}} - 2 {\overline \sigma_i}$. 
The values of $n$ and $\epsilon$ are determined from 
the plot of $log(T)$ vs $\epsilon$ for different $n$  values as shown in Fig.~\ref{FIR}. In our calculations,
we have chosen $n=120$ and the corresponding value of $\epsilon=48.7$ where the $T$ is minimum; for larger $n$ values,
the smoothing leads to a shift in the interface while for smaller $n$ values, some oscillations at the small scale remain. 
Hence, these values (which are much larger than the interplanar spacing but smaller than the simulation cell) are used 
for smoothing the fluctuations in the stress.
 
\begin{figure}[h!]
\begin{center}
\includegraphics[scale=0.28]{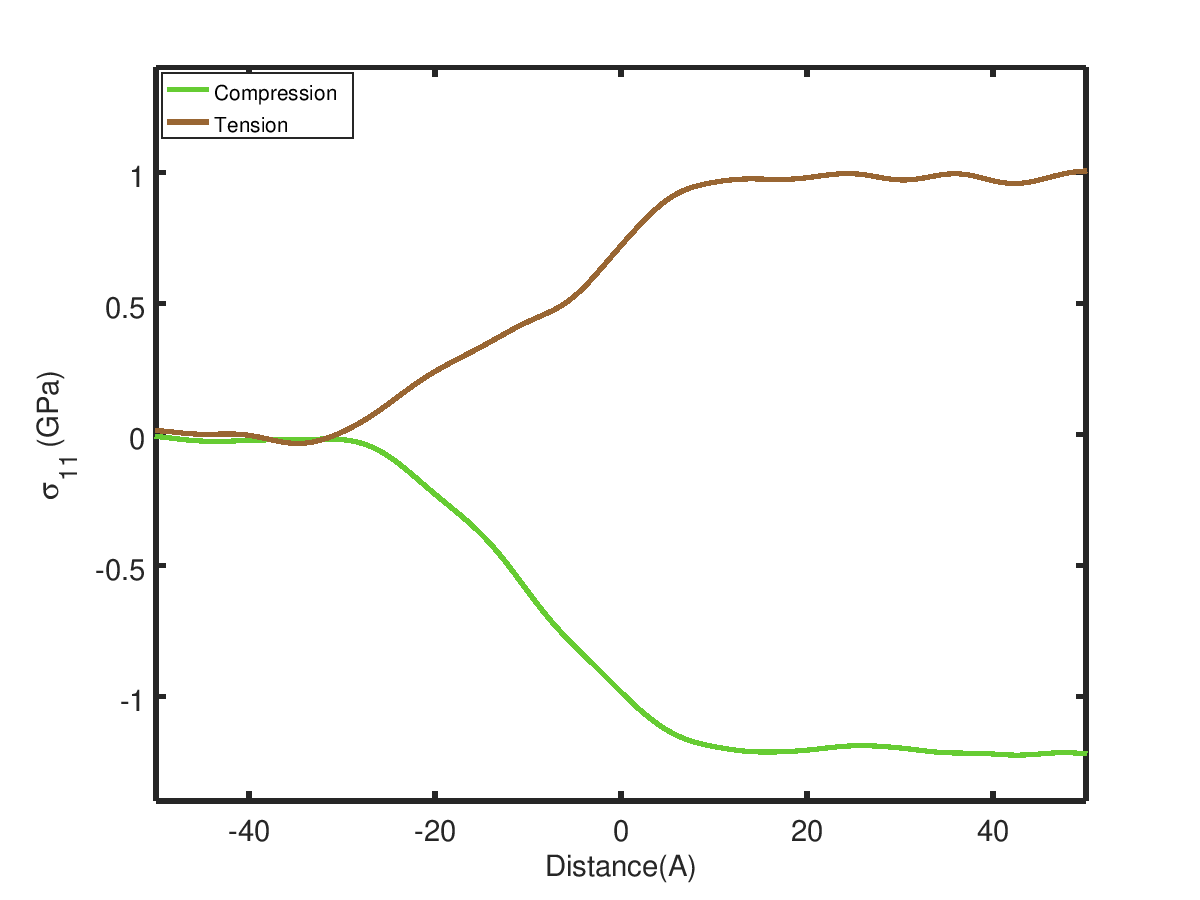}
\includegraphics[scale=0.28]{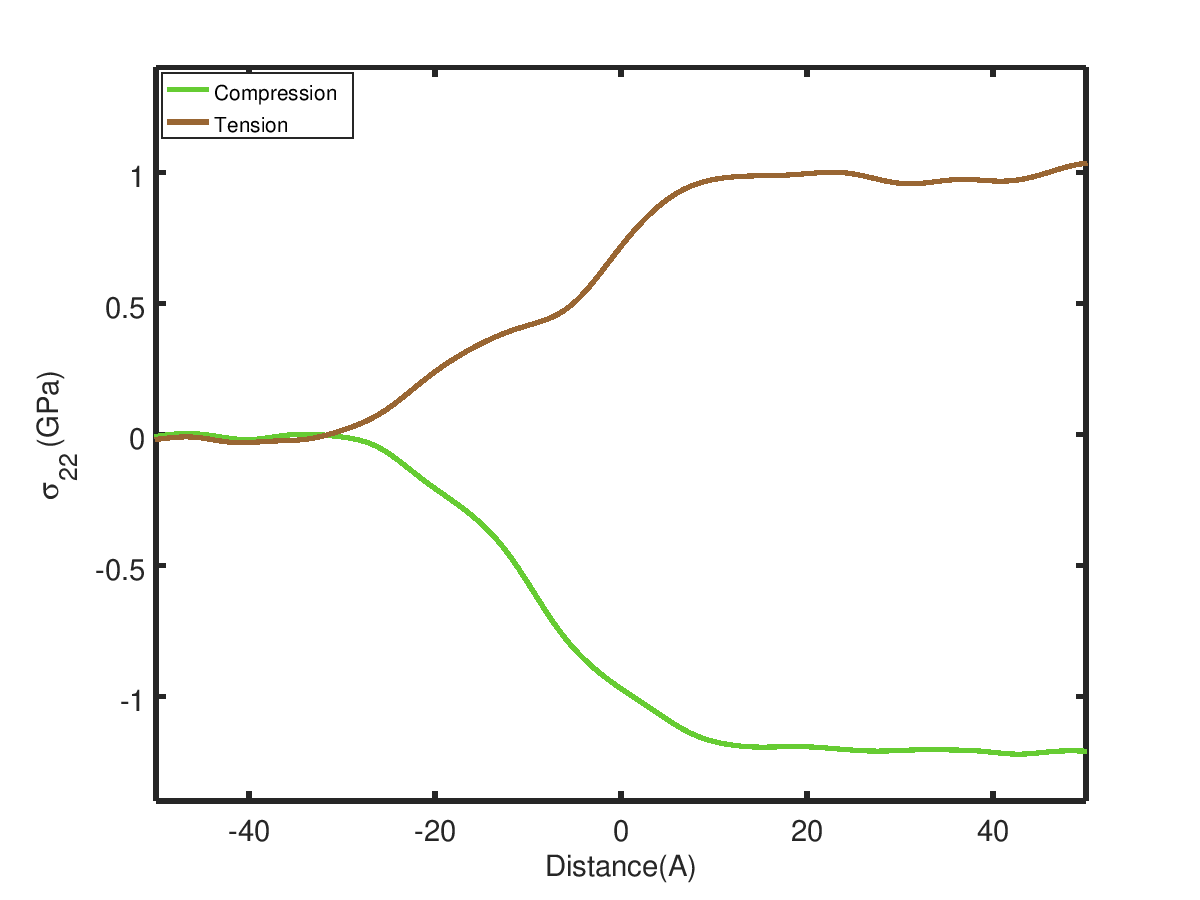}
\end{center}
\caption{FIR smoothed stress plots of $\sigma_{11}$ and $\sigma_{22}$ }\label{SmoothedStress}
\end{figure}

The smoothened stresses are shown in Fig.~\ref{SmoothedStress}. Note that the stress 
components in the solid are different owing to the elastic anisotropy~\cite{Frolov2010a}.
 
\begin{figure}[h!]
%\begin{center}
\includegraphics[scale=0.5]{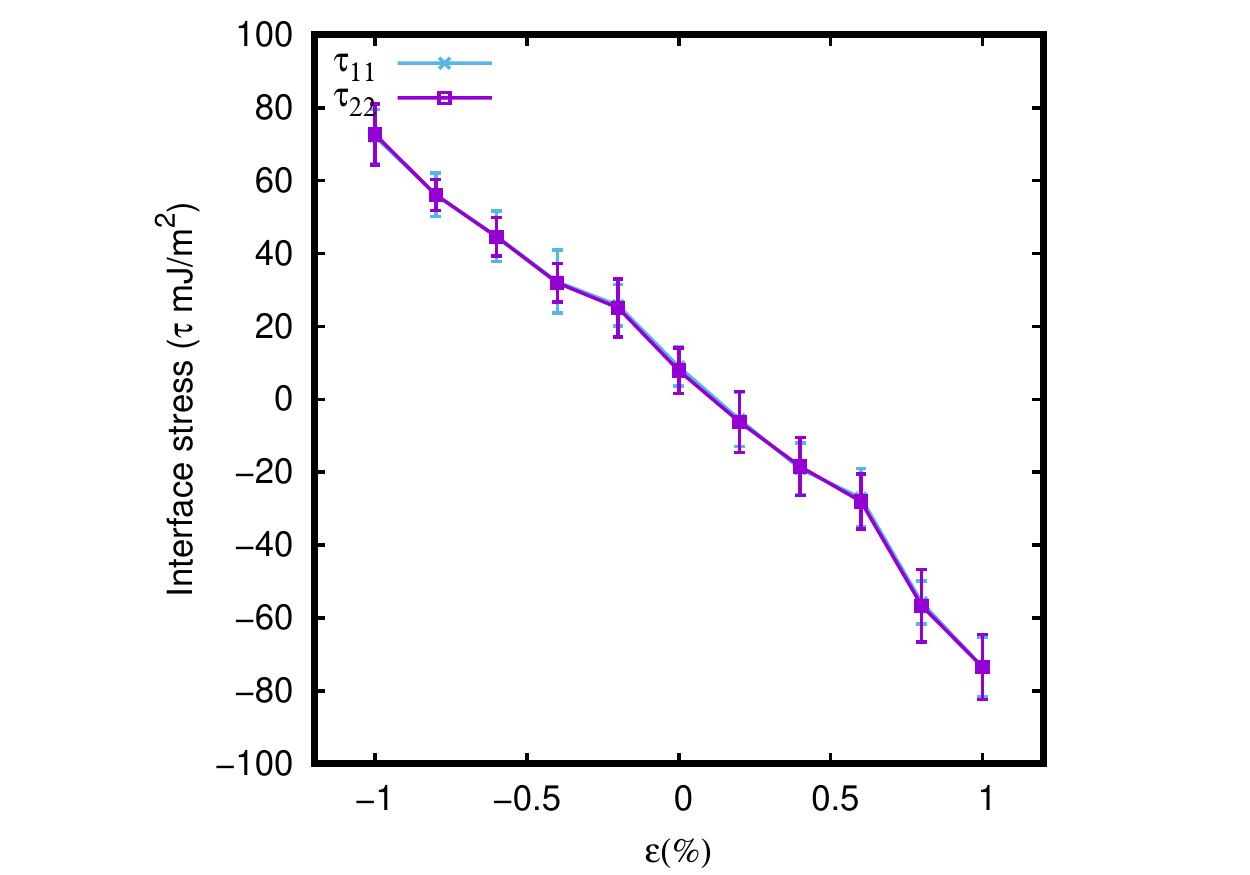}
\includegraphics[scale=0.5]{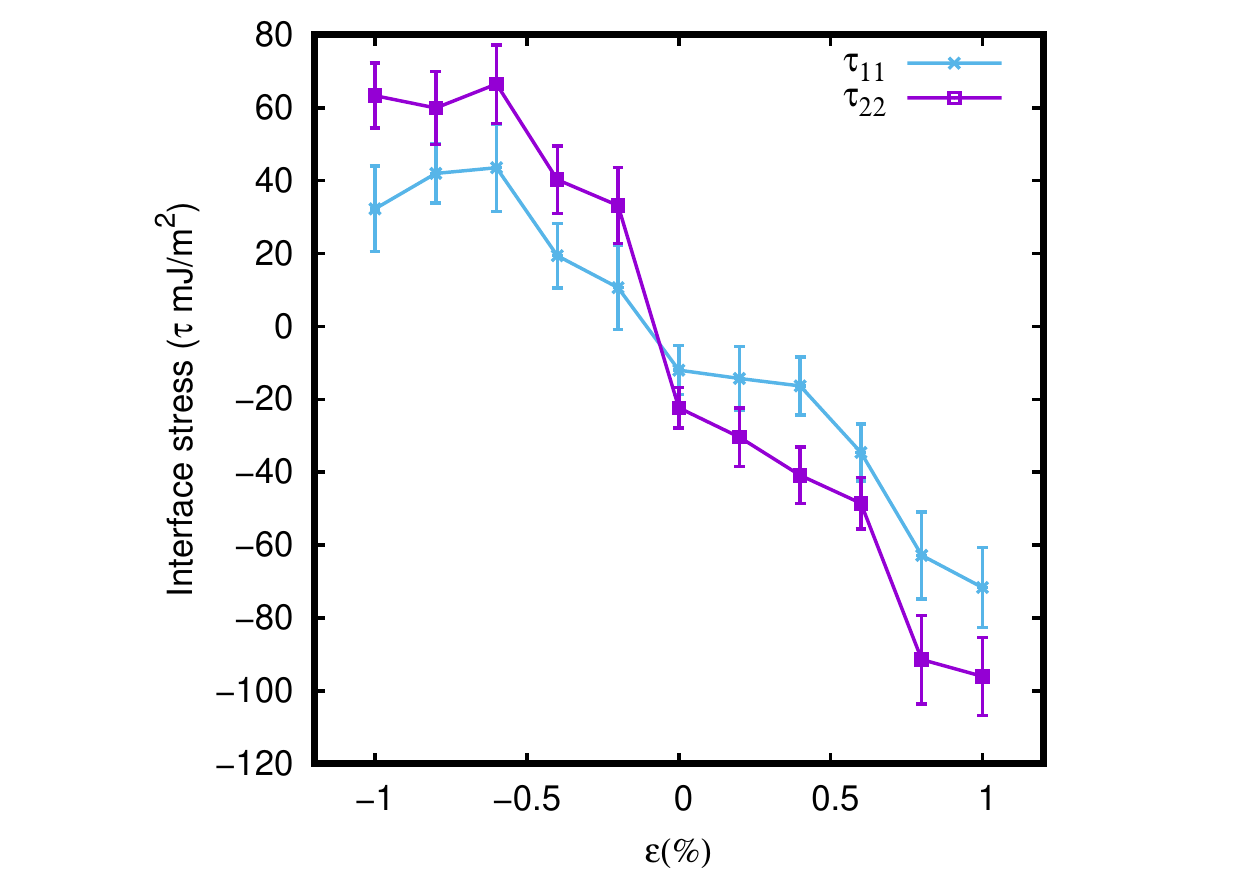}
\includegraphics[scale=0.5]{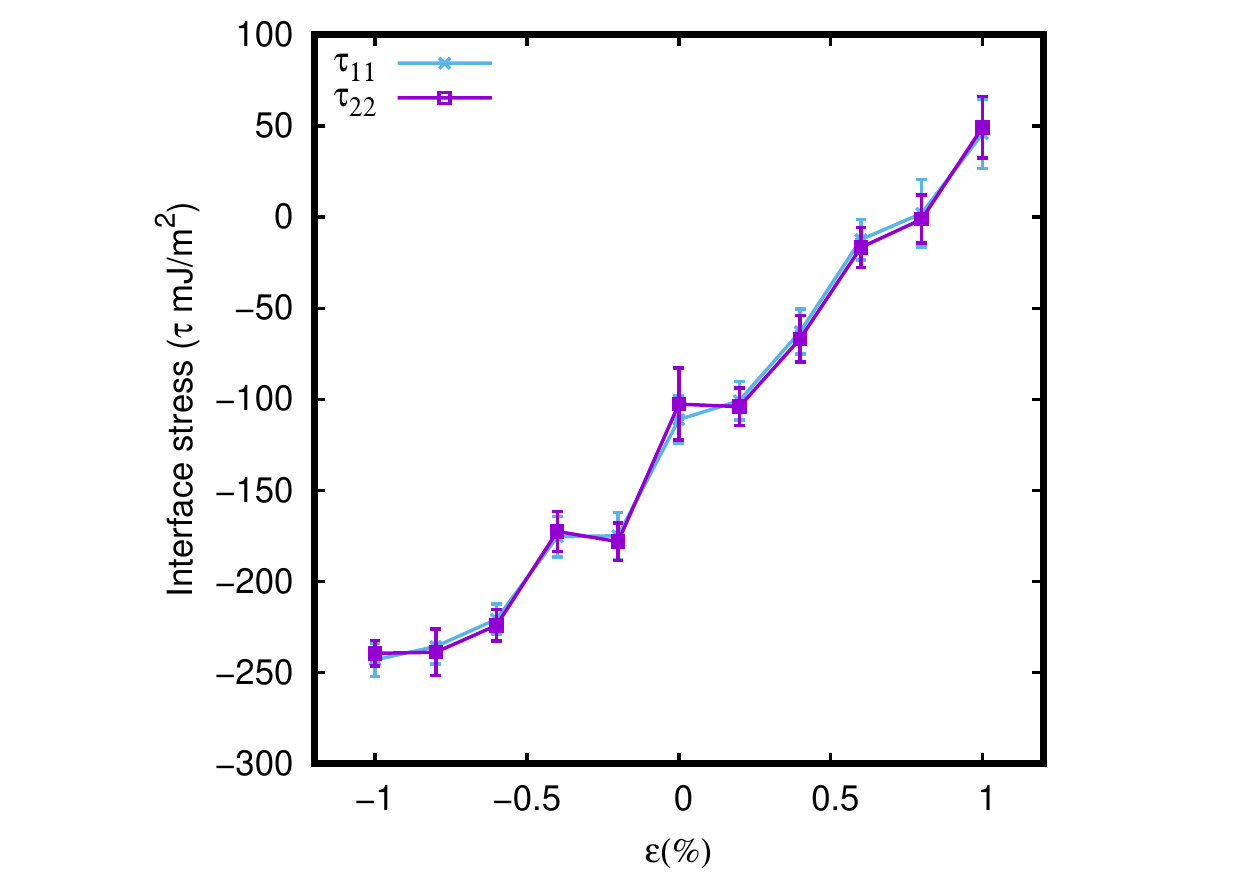}
%\end{center}
\caption{Interface stress components calculated as functions of bi-axial strain for Fe using MD 
for the interfaces orientated at (a) [001] (b) [110] (c) [111].}
\label{ist_intf}
\end{figure}

As described earlier, the solid is deformed by bi-axial strains in compression and tension, 
and the resulting interface stress is calculated for each orientation. The interface 
stress for an isothermal path having interfaces of [001], [110] and [111] orientation for the solid-melt 
interface are shown in Fig.~\ref{ist_intf}. These results indicate that the interface stress is in the range
of -250 mJ/m$^2$ to 80 mJ/m$^2$ for systems stressed to about 1\% biaxially (in compression and tension). 
Further, in all three cases, in the absence of applied strains, the interface stress is non-zero; for the 
[001] interface, this stress is tensile ($\approx 8-9$mJ/m$^2$); for [110] and [111] interfaces, 
the interface stress is compressive (and of the order of 10 and 100 mJ/m$^2$, respectively). 
The interface stress component $\tau_{11}$ and $\tau_{22}$ for 
[001] and [111] are equal as expected from the symmetry considerations and anisotropy is observed in the case 
of [110] orientation (that is, the ratio of the stresses deviates from unity). Further, in the unstressed solid, 
the anisotropy of the interface stress is considerable (0.5) and the anisotropy varies from 0.53 to 0.74. 
As seen in the case of {\em fcc} solids~\cite{Frolov2010b}, in {\em bcc} also we see that the anisotropy changes 
sign when the imposed strains change from compressive to tensile.

\newpage
\subsection{Results from PFC and Amplitude equations}

\begin{figure}[H]
\centering
\includegraphics[width=1\textwidth]{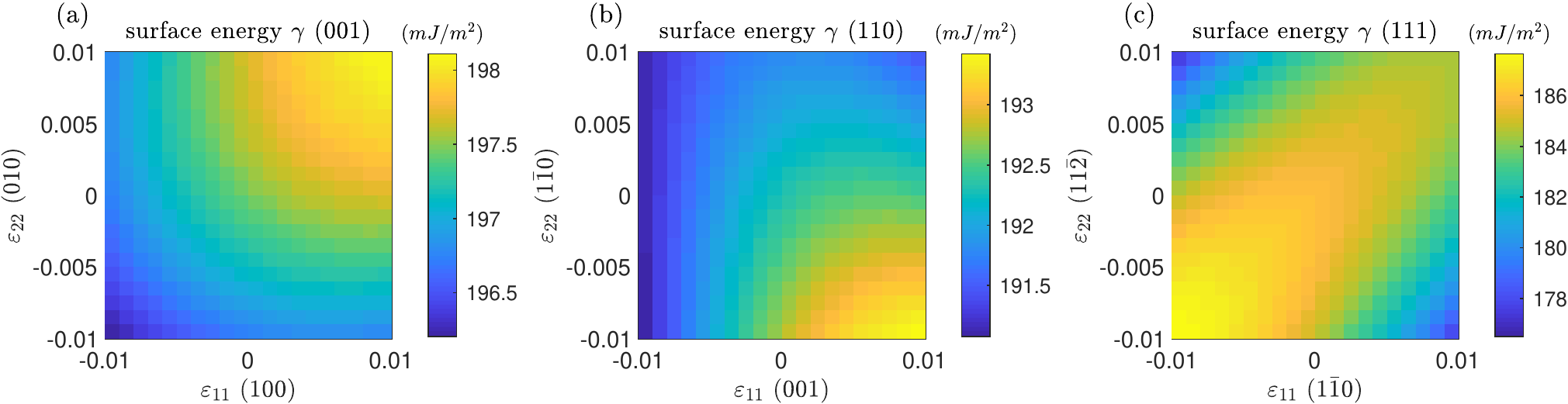}
\caption{The interfacial energy dependence on in-plane applied strains in PFC model for (a) the [100] orientation, (b) the [110] orientation and (c) the [111] orientation. The color indicates the value of interfacial energy $\gamma$ in the unit of $mJ/m^2$. For [001],[110] and [111] orientations, the applied strain $\varepsilon_{11}$ corresponds to [$100$], [$001$] and [$1\bar{1}0$] directions, respectively, while the other applied strain $\varepsilon_{22}$ corresponds to [$010$], [$1\bar{1}0$] and [$11\bar{2}$] directions, respectively.}
\label{Fig:SurfaceEnergy_PFC}
\end{figure}

\begin{figure}[H]
\centering
\includegraphics[width=1\textwidth]{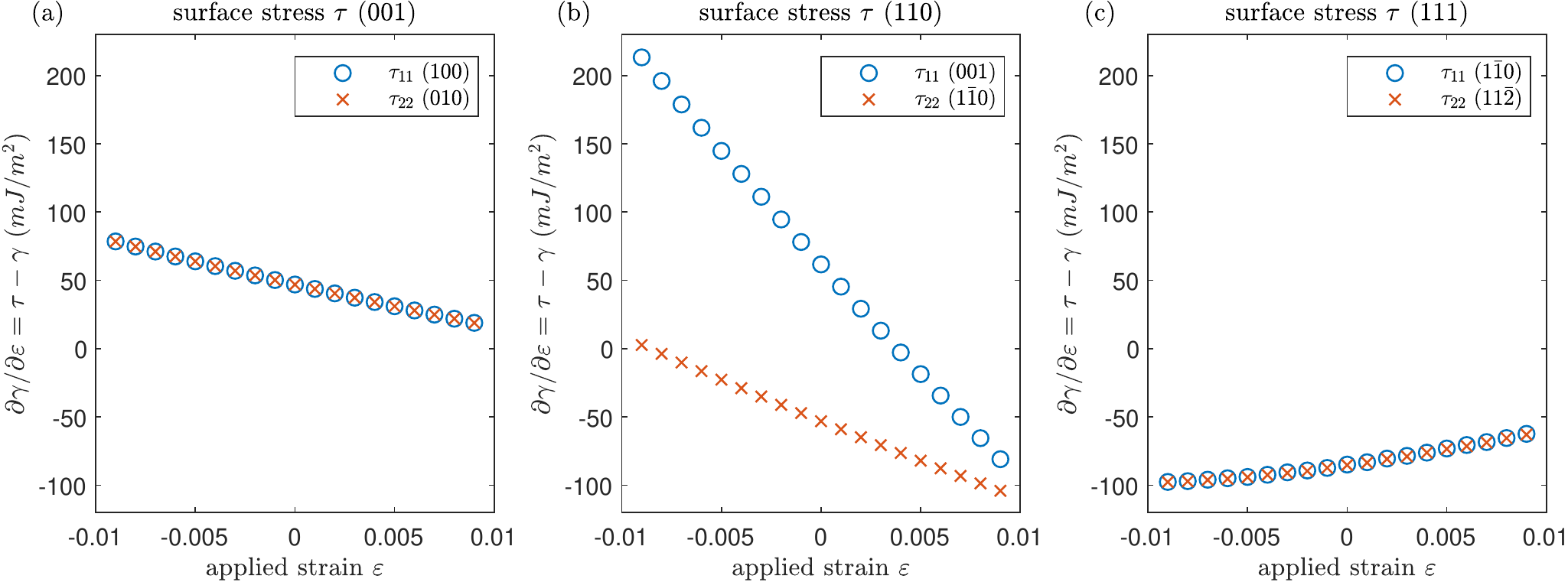}
\caption{The interfacial stress, $\tau=\partial\gamma/\partial\varepsilon_{tt}$, in PFC model for (a) the [100] orientation, (b) the [110] orientation and (c) the [111] orientation. The surface is subject to the in-plane biaxial strains, namely, $\varepsilon_{11} = \varepsilon_{22} = \varepsilon$.}
\label{Fig:SurfaceStress_PFC}
\end{figure}

The strain-dependence of interfacial energy and interfacial stress are shown in Fig.~\ref{Fig:SurfaceEnergy_PFC} and Fig.~\ref{Fig:SurfaceStress_PFC}, respectively, for PFC model under isothermal conditions for [001], [110] and [111] orientations. Note that the interfacial stress for PFC model is computed from the strain-dependent interfacial energy based on the Shuttleworth equation, $\tau_{tt} = \gamma +\partial\gamma/\partial\varepsilon_{tt}$. The simulation results are summarized in Table \ref{Table:PFC_AE_result} for the PFC model and the AE method.

\begin{table}[h!]
\begin{center}
\begin{tabular}{c | c c c | c c c}
\hline
 & $\gamma$ [100] & $\gamma$ [110] & $\gamma$ [111] & 
 $\tau_{11}$ [100] & $\tau_{11}$ / $\tau_{22}$ [110] & $\tau_{11}$ [111]\\
 & ($mJ/m^2$) & ($mJ/m^2$) & ($mJ/m^2$) &($mJ/m^2$) &($mJ/m^2$) &($mJ/m^2$)\\
\hline
PFC & $197.5$ & $192.3$ & $185.8$ & $244.4$ & $254.0$ / $139.2$ & $100.5$ \\
AE & $188.3$ & $183.4$ & $177.3$ & $219.5$ & $224.2$ / $122.3$ & $83.4$ \\
\hline
\end{tabular}
\caption{The interfacial energy $\gamma$ and interfacial stress $\tau$ for strain-free solid ($\varepsilon=0$). The values in the table are presented in the $mJ/m^2$ unit for $\gamma$ and $\tau$. Note that $\tau_{11}=\tau_{22}$ for the [100] and [111] orientations due to the lattice symmetry.}\label{Table:PFC_AE_result}
\end{center}
\end{table}

\subsection{Qualitative and quantitative comparison of results from the three methodologies}

As we see from Table.~\ref{Table:PFC_AE_result}, the values of interfacial energies and interface stresses
for PFC and the amplitude equations are in good quantitative agreement; specifically, the amplitude equation gives values
that are always less than PFC and the difference is about 4.6\% to about 17\%. 

\begin{figure}[h!]
\centering
\includegraphics[width=1\textwidth]{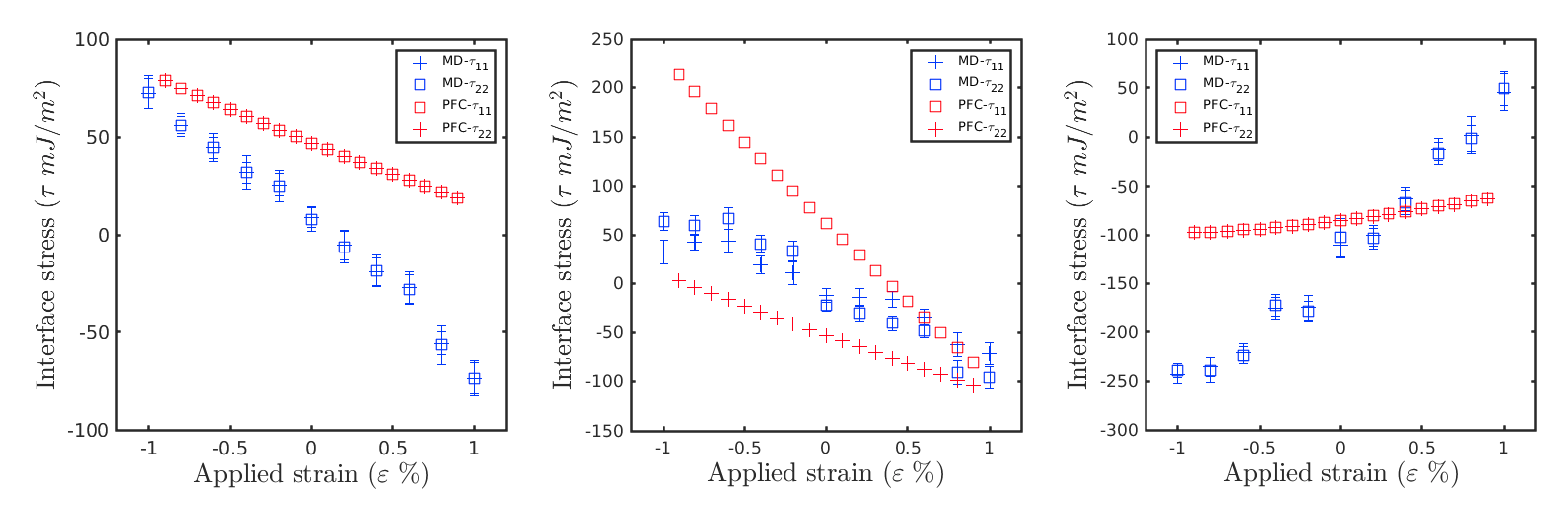}
\caption{Comparison of the interfacial stress, $\partial\gamma/\partial\varepsilon_{tt}$, obtained in MD and PFC models 
for (a) the [100] orientation, (b) the [110] orientation and (c) the [111] orientation. The surface is subject to the in-plane biaxial strains, 
namely, $\varepsilon_{11} = \varepsilon_{22} = \varepsilon$.}
\label{ComparisonFig}
\end{figure}

On the other hand, as can be seen from Fig.~\ref{ComparisonFig},
the MD and PFC/AE are qualitatively similar in most cases; for example, in all cases, 
the trends are the same in both MD and PFE/AE-- namely, for both $(100)$ and $(110)$, 
the plots show a negative slope while for $(111)$ the plots show a positive slope. However, they are different 
quantitatively; the range and the nature
of the stress (compressive or tensile) is different -- even though the order of magnitude is the same, namely,
about a couple of hundred mJ/m$^2$. 

The quantitative differences in the cases of $(110)$ and $(111)$ are the most striking. 
In MD, for $(110)$, the anisotropy is less for unstressed solid; and, it varies from 0.54 and 0.74 and the nature
of anisotropy changes with the change in the nature of imposed strain. In PFC/AE, the tensile imposed strain
of about 1\% has the least anisotropy (of 0.77) and the compressive imposed strain of about 1\% has the highest anisotropy
(of 80.38). Interestingly, for the unstressed solid, the anisotropy is very high and is (1.16). Similarly, the differences
in slope of the interface stress obtained using MD and that obtained using PFC are very different for the $(111)$ system. 

In the case of MD, we have used EAM potentials; however, the PFC model employed in this study uses a single parameter  
to model the system. As shown in previous study, the PFC parameter sets the interface width, and the PFC is shown to 
predict reasonable interface energy and its anisotropy. However, due to the simplicity of the PFC method, the strain-stress 
relation of solids are uniquely determined by this parameter at the same time, which may not be realistic.    
We believe that the discrepancies arise from this difference in the two models, and if the PFC/AE model is modified to capture
the elastic moduli more accurately, the quantitative agreement between MD and PFC will improve. Having said that,
broadly, for qualitative understanding the anisotropy, the information captures by the PFC model with one parameter
(unlike MD which uses EAM potential), is remarkable.

\section{Conclusions} \label{Section5}

We have used both MD and PFC to study the anisotropy in interface stress at the {\em bcc}-iron crystal-melt
interface. Our results show that the interface stress is anisotropic; this has implications for the equilibrium
and growth morphologies of iron particles at the nano-scale. Our study also indicates that with a single parameter,
PFC captures the interface stress anisotropy qualitatively and gives stresses at the same order as MD. We also
show that the amplitude equations based on PFC can also be used to calculate the interface stresses and the values
thus obtained are in good agreement with PFC. Finally, our results show that even in the case of weak anisotropy
in interfacial energy, the interface stress anisotropy can be strong.

\section*{Acknowledgements}

S. Kumar and M. P. Gururajan thank (i) C-DAC, Pune, (ii) IIT-Bombay, and (iii) the DST-FIST facility at the 
Department of Metallurgical  Engineering and Materials Science, IIT-Bombay for the computational facilities 
and G Kamalakshi for useful discussions. M.-W. Liu and K.-A. Wu gratefully acknowledge the support of the Ministry 
of Science and Technology, Taiwan (Grant No. MOST 109-2112-M-007-005-), and the support from the National Center 
for Theoretical Sciences, Taiwan.

\bibliography{mybibfile}

\begin{thebibliography}{10}
\expandafter\ifx\csname url\endcsname\relax
  \def\url#1{\texttt{#1}}\fi
\expandafter\ifx\csname urlprefix\endcsname\relax\def\urlprefix{URL }\fi
\expandafter\ifx\csname href\endcsname\relax
  \def\href#1#2{#2} \def\path#1{#1}\fi

\bibitem{PorterEasterling}
D.~A. Porter, K.~E. Easterling, M.~Y. Sherif, Phase Transformations in Metals
  and Alloys, 3rd Edition, CRC Press, 2009.

\bibitem{Turnbull1950}
D.~Turnbull, R.~E. Cech, {Microscopic observation of the solidification of
  small metal droplets}, Journal of Applied Physics 21~(8) (1950) 804--810.
\newblock \href {https://doi.org/10.1063/1.1699763}
  {\path{doi:10.1063/1.1699763}}.

\bibitem{AstaEtAl2009}
Y.~Mishin, M.~Asta, J.~Li,
  \href{http://dx.doi.org/10.1016/j.actamat.2009.10.049}{Atomistic modeling of
  interfaces and their impact on microstructure and properties}, Acta
  Materialia 58~(4) (2010) 1117--1151.
\newblock \href {https://doi.org/10.1016/j.actamat.2009.10.049}
  {\path{doi:10.1016/j.actamat.2009.10.049}}.
\newline\urlprefix\url{http://dx.doi.org/10.1016/j.actamat.2009.10.049}

\bibitem{LiuEtAl2001}
S.~Liu, R.~E. Napolitano, R.~Trivedi, Measurement of anisotropy of crystal-melt
  interfacial energy for a binary {Al-Cu} alloy, Acta Materialia 49~(20) (2001)
  4271--4276.
\newblock \href {https://doi.org/10.1016/S1359-6454(01)00306-8}
  {\path{doi:10.1016/S1359-6454(01)00306-8}}.

\bibitem{BackofenVoigt2009}
R.~Backofen, A.~Voigt, Solid-liquid interfacial energies and equilibrium shapes
  of nanocrystals, Journal of Physics Condensed Matter 21~(46) (2009).
\newblock \href {https://doi.org/10.1088/0953-8984/21/46/464109}
  {\path{doi:10.1088/0953-8984/21/46/464109}}.

\bibitem{Sekerka2005}
R.~F. Sekerka, Equilibrium and growth shapes of crystals: How do they differ
  and why should we care?, Crystal Research and Technology 40~(4-5) (2005)
  291--306.
\newblock \href {https://doi.org/10.1002/crat.200410342}
  {\path{doi:10.1002/crat.200410342}}.

\bibitem{NapolitanoEtAl2002}
R.~E. Napolitano, S.~Liu, R.~Trivedi, Experimental measurement of anisotropy in
  crystal-melt interfacial energy, Interface Science 10~(2-3) (2002) 217--232.
\newblock \href {https://doi.org/10.1023/A:1015884415896}
  {\path{doi:10.1023/A:1015884415896}}.

\bibitem{Davidchack2000}
R.~L. Davidchack, B.~B. Laird,
  \href{https://link.aps.org/doi/10.1103/PhysRevLett.85.4751}{Direct
  calculation of the hard-sphere crystal $/$melt interfacial free energy},
  Phys. Rev. Lett. 85 (2000) 4751--4754.
\newblock \href {https://doi.org/10.1103/PhysRevLett.85.4751}
  {\path{doi:10.1103/PhysRevLett.85.4751}}.
\newline\urlprefix\url{https://link.aps.org/doi/10.1103/PhysRevLett.85.4751}

\bibitem{HoytEtAl2001}
J.~J. Hoyt, M.~Asta, A.~Karma,
  \href{https://link.aps.org/doi/10.1103/PhysRevLett.86.5530}{Method for
  computing the anisotropy of the solid-liquid interfacial free energy}, Phys.
  Rev. Lett. 86 (2001) 5530--5533.
\newblock \href {https://doi.org/10.1103/PhysRevLett.86.5530}
  {\path{doi:10.1103/PhysRevLett.86.5530}}.
\newline\urlprefix\url{https://link.aps.org/doi/10.1103/PhysRevLett.86.5530}

\bibitem{HoytEtAl2004}
J.~J. Hoyt, M.~Asta, T.~Haxhimali, A.~Karma, R.~E. Napolitano, R.~Trivedi,
  B.~B. Laird, J.~R. Morris, Crystal-melt interfaces and solidification
  morphologies in metals and alloys, MRS Bulletin 29~(12) (2004) 935--939.
\newblock \href {https://doi.org/10.1557/mrs2004.263}
  {\path{doi:10.1557/mrs2004.263}}.

\bibitem{WuKarma2007}
K.-A. Wu, A.~Karma,
  \href{https://link.aps.org/doi/10.1103/PhysRevB.76.184107}{Phase-field
  crystal modeling of equilibrium bcc-liquid interfaces}, Phys. Rev. B 76
  (2007) 184107.
\newblock \href {https://doi.org/10.1103/PhysRevB.76.184107}
  {\path{doi:10.1103/PhysRevB.76.184107}}.
\newline\urlprefix\url{https://link.aps.org/doi/10.1103/PhysRevB.76.184107}

\bibitem{Jaatinen2009}
A.~Jaatinen, C.~V. Achim, K.~R. Elder, T.~Ala-Nissila,
  \href{https://link.aps.org/doi/10.1103/PhysRevE.80.031602}{Thermodynamics of
  bcc metals in phase-field-crystal models}, Phys. Rev. E 80 (2009) 031602.
\newblock \href {https://doi.org/10.1103/PhysRevE.80.031602}
  {\path{doi:10.1103/PhysRevE.80.031602}}.
\newline\urlprefix\url{https://link.aps.org/doi/10.1103/PhysRevE.80.031602}

\bibitem{Toth2014}
G.~I. T{\'{o}}th, N.~Provatas,
  \href{https://link.aps.org/doi/10.1103/PhysRevB.90.104101}{Advanced
  ginzburg-landau theory of freezing: A density-functional approach}, Phys.
  Rev. B 90 (2014) 104101.
\newblock \href {https://doi.org/10.1103/PhysRevB.90.104101}
  {\path{doi:10.1103/PhysRevB.90.104101}}.
\newline\urlprefix\url{https://link.aps.org/doi/10.1103/PhysRevB.90.104101}

\bibitem{LeoSekerka}
P.~H. Leo, R.~F. Sekerka, \href{<Go to ISI>://WOS:A1989CG88000001}{The effect
  of surface stress on crystal melt and crystal crystal equilibrium}, Acta
  Metallurgica 37~(12) (1989) 3119--3138.
\newline\urlprefix\url{<Go to ISI>://WOS:A1989CG88000001}

\bibitem{Herring}
C.~Herring, The use of classical macroscopic concepts in surface-energy
  problems, in: R.~Gomer, C.~S. Smith (Eds.), Structure and Properties of Solid
  Surfaces; proceedings of a conference arranged by the National Research
  Council and held in September, 1952, in Lake Geneva, Wisconsin, USA, Chicago
  Press, Chicago, 1953, Ch.~1, pp. 5--81.

\bibitem{Shuttleworth}
R.~Shuttleworth, \href{http://stacks.iop.org/0370-1298/63/i=5/a=302}{The
  surface tension of solids}, Proceedings of the Physical Society. Section A
  63~(5) (1950) 444.
\newline\urlprefix\url{http://stacks.iop.org/0370-1298/63/i=5/a=302}

\bibitem{JohnsonAlexander}
W.~C. Johnson, J.~I.~D. Alexander, {Interfacial conditions for thermomechanical
  equilibrium in two-phase crystals}, Journal of Applied Physics 59~(8) (1986)
  2735--2746.
\newblock \href {https://doi.org/10.1063/1.336982}
  {\path{doi:10.1063/1.336982}}.

\bibitem{Eremeyev2016}
V.~A. Eremeyev, On effective properties of materials at the nano- and
  microscales considering surface effects, Acta Mechanica 227~(1) (2016)
  29--42.
\newblock \href {https://doi.org/10.1007/s00707-015-1427-y}
  {\path{doi:10.1007/s00707-015-1427-y}}.

\bibitem{MomeniLevitas2016}
K.~Momeni, V.~I. Levitas, A phase-field approach to nonequilibrium phase
  transformations in elastic solids: Via an intermediate phase (melt) allowing
  for interface stresses, Physical Chemistry Chemical Physics 18~(17) (2016)
  12183--12203.
\newblock \href {https://doi.org/10.1039/c6cp00943c}
  {\path{doi:10.1039/c6cp00943c}}.

\bibitem{Frolov2009}
T.~Frolov, Y.~Mishin, Temperature dependence of the surface free energy and
  surface stress: An atomistic calculation for cu(110), Physical Review B
  79~(4) (2009).
\newblock \href {https://doi.org/10.1103/PhysRevB.79.045430}
  {\path{doi:10.1103/PhysRevB.79.045430}}.

\bibitem{Frolov2010}
T.~Frolov, Y.~Mishin, \href{<Go to ISI>://WOS:000282130700004}{Orientation
  dependence of the solid-liquid interface stress: atomistic calculations for
  copper}, Modelling and Simulation in Materials Science and Engineering 18~(7)
  (2010).
\newline\urlprefix\url{<Go to ISI>://WOS:000282130700004}

\bibitem{Frolov2010b}
T.~Frolov, Y.~Mishin, Effect of nonhydrostatic stresses on solid-fluid
  equilibrium. {I}. bulk thermodynamics, Physical Review B - Condensed Matter
  and Materials Physics 82~(17) (2010) 1--14.
\newblock \href {https://doi.org/10.1103/PhysRevB.82.174113}
  {\path{doi:10.1103/PhysRevB.82.174113}}.

\bibitem{Frolov2010a}
T.~Frolov, Y.~Mishin, Effect of nonhydrostatic stresses on solid-fluid
  equilibrium. {II}. interface thermodynamics, Physical Review B 82~(17)
  (2010).
\newblock \href {https://doi.org/10.1103/PhysRevB.82.174114}
  {\path{doi:10.1103/PhysRevB.82.174114}}.

\bibitem{LiuWu2017}
M.~W. Liu, K.~A. Wu, {Investigation of surface/bulk stresses of nanoparticles
  with diffusive interfaces using the phase field crystal model}, Physical
  Review B 96~(21) (2017).
\newblock \href {https://doi.org/10.1103/PhysRevB.96.214106}
  {\path{doi:10.1103/PhysRevB.96.214106}}.

\bibitem{Wu2006}
K.-A. Wu, A.~Karma, J.~J. Hoyt, M.~Asta,
  \href{https://link.aps.org/doi/10.1103/PhysRevB.73.094101}{Ginzburg-landau
  theory of crystalline anisotropy for bcc-liquid interfaces}, Phys. Rev. B 73
  (2006) 094101.
\newblock \href {https://doi.org/10.1103/PhysRevB.73.094101}
  {\path{doi:10.1103/PhysRevB.73.094101}}.
\newline\urlprefix\url{https://link.aps.org/doi/10.1103/PhysRevB.73.094101}

\bibitem{Mendelev2003}
M.~I. Mendelev, S.~Han, D.~J. Srolovitz, G.~J. Ackland, D.~Y. Sun, M.~Asta,
  Development of new interatomic potentials appropriate for crystalline and
  liquid iron, Philosophical Magazine 83~(35) (2003) 3977--3994.
\newblock \href {https://doi.org/10.1080/14786430310001613264}
  {\path{doi:10.1080/14786430310001613264}}.

\bibitem{Morris2002}
J.~R. Morris, X.~Song, The melting lines of model systems calculated from
  coexistence simulations, Journal of Chemical Physics 116~(21) (2002)
  9352--9358.
\newblock \href {https://doi.org/10.1063/1.1474581}
  {\path{doi:10.1063/1.1474581}}.

\bibitem{Liu2013}
J.~Liu, R.~L. Davidchack, H.~B. Dong, Molecular dynamics calculation of
  solid-liquid interfacial free energy and its anisotropy during iron
  solidification, Computational Materials Science 74 (2013) 92--100.
\newblock \href {https://doi.org/10.1016/j.commatsci.2013.03.018}
  {\path{doi:10.1016/j.commatsci.2013.03.018}}.

\bibitem{Ramakrishnan2017}
R.~Ramakrishnan, R.~Sankarasubramanian, Crystal-melt kinetic coefficients of
  {$Ni_3Al$}, Acta Materialia 127 (2017) 25--32.
\newblock \href {https://doi.org/10.1016/j.actamat.2017.01.009}
  {\path{doi:10.1016/j.actamat.2017.01.009}}.

\bibitem{LAMMPSManual}
S.~Plimpton, Fast parallel algorithms for short-range molecular dynamics,
  Journal of Computational Physics 117~(1) (1995) 1--19.
\newblock \href {https://doi.org/10.1006/jcph.1995.1039}
  {\path{doi:10.1006/jcph.1995.1039}}.

\bibitem{Rayne1961}
J.~A. Rayne, B.~S. Chandrasekhar, {Elastic constants of iron from 4.2 to
  300°K}, Physical Review 122~(6) (1961) 1714--1716.
\newblock \href {https://doi.org/10.1103/PhysRev.122.1714}
  {\path{doi:10.1103/PhysRev.122.1714}}.

\bibitem{Etesami2018}
S.~A. Etesami, E.~Asadi,
  \href{https://doi.org/10.1016/j.jpcs.2017.09.001}{{Molecular dynamics for
  near melting temperatures simulations of metals using modified embedded-atom
  method}}, Journal of Physics and Chemistry of Solids 112~(July 2017) (2018)
  61--72.
\newblock \href {https://doi.org/10.1016/j.jpcs.2017.09.001}
  {\path{doi:10.1016/j.jpcs.2017.09.001}}.
\newline\urlprefix\url{https://doi.org/10.1016/j.jpcs.2017.09.001}

\bibitem{Davidchack2006}
R.~L. Davidchack, J.~R. Morris, B.~B. Laird, The anisotropic hard-sphere
  crystal-melt interfacial free energy from fluctuations, Journal of Chemical
  Physics 125~(9) (2006).
\newblock \href {https://doi.org/10.1063/1.2338303}
  {\path{doi:10.1063/1.2338303}}.

\bibitem{Elder2002}
K.~R. Elder, M.~Katakowski, M.~Haataja, M.~Grant, {Modeling elasticity in
  crystal growth}, Physical Review Letters 88~(24) (2002) 2457011--2457014.
\newblock \href {http://arxiv.org/abs/0107381} {\path{arXiv:0107381}}, \href
  {https://doi.org/10.1103/PhysRevLett.88.245701}
  {\path{doi:10.1103/PhysRevLett.88.245701}}.

\bibitem{Elder2004}
K.~R. Elder, M.~Grant,
  \href{http://link.aps.org/doi/10.1103/PhysRevE.70.051605}{Modeling elastic
  and plastic deformations in nonequilibrium processing using phase field
  crystals}, Phys. Rev. E 70 (2004) 051605.
\newblock \href {https://doi.org/10.1103/PhysRevE.70.051605}
  {\path{doi:10.1103/PhysRevE.70.051605}}.
\newline\urlprefix\url{http://link.aps.org/doi/10.1103/PhysRevE.70.051605}

\bibitem{Pisutha-Arnond2013}
N.~Pisutha-Arnond, V.~W.~L. Chan, K.~R. Elder, K.~Thornton,
  \href{http://link.aps.org/doi/10.1103/PhysRevB.87.014103}{Calculations of
  isothermal elastic constants in the phase-field crystal model}, Phys. Rev. B
  87 (2013) 014103.
\newblock \href {https://doi.org/10.1103/PhysRevB.87.014103}
  {\path{doi:10.1103/PhysRevB.87.014103}}.
\newline\urlprefix\url{http://link.aps.org/doi/10.1103/PhysRevB.87.014103}

\bibitem{Heinonen2014}
V.~Heinonen, C.~V. Achim, K.~R. Elder, S.~Buyukdagli, T.~Ala-Nissila,
  \href{http://link.aps.org/doi/10.1103/PhysRevE.89.032411}{Phase-field-crystal
  models and mechanical equilibrium}, Phys. Rev. E 89 (2014) 032411.
\newblock \href {https://doi.org/10.1103/PhysRevE.89.032411}
  {\path{doi:10.1103/PhysRevE.89.032411}}.
\newline\urlprefix\url{http://link.aps.org/doi/10.1103/PhysRevE.89.032411}

\bibitem{Wu2009}
K.-A. Wu, P.~W. Voorhees,
  \href{https://link.aps.org/doi/10.1103/PhysRevB.80.125408}{Stress-induced
  morphological instabilities at the nanoscale examined using the phase field
  crystal approach}, Phys. Rev. B 80 (2009) 125408.
\newblock \href {https://doi.org/10.1103/PhysRevB.80.125408}
  {\path{doi:10.1103/PhysRevB.80.125408}}.
\newline\urlprefix\url{https://link.aps.org/doi/10.1103/PhysRevB.80.125408}

\bibitem{Yu2011}
Y.~M. Yu, R.~Backofen, A.~Voigt,
  \href{http://dx.doi.org/10.1016/j.jcrysgro.2010.08.047}{Morphological
  instability of heteroepitaxial growth on vicinal substrates: A phase-field
  crystal study}, Journal of Crystal Growth 318~(1) (2011) 18--22.
\newblock \href {https://doi.org/10.1016/j.jcrysgro.2010.08.047}
  {\path{doi:10.1016/j.jcrysgro.2010.08.047}}.
\newline\urlprefix\url{http://dx.doi.org/10.1016/j.jcrysgro.2010.08.047}

\bibitem{Choudhary2014}
M.~A. Choudhary, J.~Kundin, H.~Emmerich,
  \href{http://dx.doi.org/10.1016/j.commatsci.2013.11.030}{Misfit and
  dislocation nucleation during heteroepitaxial growth}, Computational
  Materials Science 83 (2014) 481--487.
\newblock \href {https://doi.org/10.1016/j.commatsci.2013.11.030}
  {\path{doi:10.1016/j.commatsci.2013.11.030}}.
\newline\urlprefix\url{http://dx.doi.org/10.1016/j.commatsci.2013.11.030}

\bibitem{Podmaniczky2017}
F.~Podmaniczky, G.~I. T{\'{o}}th, G.~Tegze, T.~Pusztai, L.~Gr{\'{a}}n{\'{a}}sy,
  \href{http://dx.doi.org/10.1016/j.jcrysgro.2016.06.056}{{Phase-field crystal
  modeling of heteroepitaxy and exotic modes of crystal nucleation}}, Journal
  of Crystal Growth 457 (2017) 24--31.
\newblock \href {https://doi.org/10.1016/j.jcrysgro.2016.06.056}
  {\path{doi:10.1016/j.jcrysgro.2016.06.056}}.
\newline\urlprefix\url{http://dx.doi.org/10.1016/j.jcrysgro.2016.06.056}

\bibitem{Backofen2010}
R.~Backofen, A.~Voigt, A phase-field-crystal approach to critical nuclei,
  Journal of Physics Condensed Matter 22~(36) (2010).
\newblock \href {https://doi.org/10.1088/0953-8984/22/36/364104}
  {\path{doi:10.1088/0953-8984/22/36/364104}}.

\bibitem{Toth2010}
G.~T{\'{o}}th, G.~Tegze, T.~Pusztai, G.~T{\'{o}}th, L.~Gr{\'{a}}n{\'{a}}sy,
  Polymorphism, crystal nucleation and growth in the phase-field crystal model
  in {2D} and {3D}, Journal of Physics Condensed Matter 22~(36) (2010).
\newblock \href {https://doi.org/10.1088/0953-8984/22/36/364101}
  {\path{doi:10.1088/0953-8984/22/36/364101}}.

\bibitem{Tang2014}
S.~Tang, Y.~M. Yu, J.~Wang, J.~Li, Z.~Wang, Y.~Guo, Y.~Zhou,
  Phase-field-crystal simulation of nonequilibrium crystal growth, Physical
  Review E - Statistical, Nonlinear, and Soft Matter Physics 89~(1) (2014)
  1--6.
\newblock \href {https://doi.org/10.1103/PhysRevE.89.012405}
  {\path{doi:10.1103/PhysRevE.89.012405}}.

\bibitem{Elder2007}
K.~R. Elder, N.~Provatas, J.~Berry, P.~Stefanovic, M.~Grant, Phase-field
  crystal modeling and classical density functional theory of freezing,
  Physical Review B - Condensed Matter and Materials Physics 75~(6) (2007)
  1--14.
\newblock \href {https://doi.org/10.1103/PhysRevB.75.064107}
  {\path{doi:10.1103/PhysRevB.75.064107}}.

\bibitem{Wu2010}
K.-A. Wu, A.~Adland, A.~Karma,
  \href{https://link.aps.org/doi/10.1103/PhysRevE.81.061601}{Phase-field-crystal
  model for fcc ordering}, Phys. Rev. E 81 (2010) 061601.
\newblock \href {https://doi.org/10.1103/PhysRevE.81.061601}
  {\path{doi:10.1103/PhysRevE.81.061601}}.
\newline\urlprefix\url{https://link.aps.org/doi/10.1103/PhysRevE.81.061601}

\bibitem{Greenwood2010}
M.~Greenwood, N.~Provatas, J.~Rottler,
  \href{https://www.ncbi.nlm.nih.gov/pubmed/20867862}{Free energy functionals
  for efficient phase field crystal modeling of structural phase
  transformations}, Phys Rev Lett 105~(4) (2010) 045702.
\newblock \href {https://doi.org/10.1103/PhysRevLett.105.045702}
  {\path{doi:10.1103/PhysRevLett.105.045702}}.
\newline\urlprefix\url{https://www.ncbi.nlm.nih.gov/pubmed/20867862}

\bibitem{Greenwood2011}
M.~Greenwood, J.~Rottler, N.~Provatas,
  \href{https://www.ncbi.nlm.nih.gov/pubmed/21517507}{Phase-field-crystal
  methodology for modeling of structural transformations}, Phys Rev E Stat
  Nonlin Soft Matter Phys 83~(3 Pt 1) (2011) 031601.
\newblock \href {https://doi.org/10.1103/PhysRevE.83.031601}
  {\path{doi:10.1103/PhysRevE.83.031601}}.
\newline\urlprefix\url{https://www.ncbi.nlm.nih.gov/pubmed/21517507}

\bibitem{Chan2009}
P.~Y. Chan, N.~Goldenfeld,
  \href{http://link.aps.org/doi/10.1103/PhysRevE.80.065105}{Nonlinear
  elasticity of the phase-field crystal model from the renormalization group},
  Phys. Rev. E 80 (2009) 065105.
\newblock \href {https://doi.org/10.1103/PhysRevE.80.065105}
  {\path{doi:10.1103/PhysRevE.80.065105}}.
\newline\urlprefix\url{http://link.aps.org/doi/10.1103/PhysRevE.80.065105}

\bibitem{Elder2010}
K.~R. Elder, Z.~F. Huang, N.~Provatas,
  \href{https://www.ncbi.nlm.nih.gov/pubmed/20365379}{Amplitude expansion of
  the binary phase-field-crystal model}, Phys Rev E Stat Nonlin Soft Matter
  Phys 81~(1 Pt 1) (2010) 011602.
\newblock \href {https://doi.org/10.1103/PhysRevE.81.011602}
  {\path{doi:10.1103/PhysRevE.81.011602}}.
\newline\urlprefix\url{https://www.ncbi.nlm.nih.gov/pubmed/20365379}

\bibitem{Davidchack1998}
R.~L. Davidchack, B.~B. Laird, Simulation of the hard-sphere crystal–melt
  interface, The Journal of Chemical Physics 108~(22) (1998) 9452--9462.
\newblock \href {https://doi.org/10.1063/1.476396}
  {\path{doi:10.1063/1.476396}}.

\bibitem{yunfei2012}
Y.~Bai, Molecular dynamics simulation study of solid-liquid interface
  properties of {HCP} magnesium, Master's thesis, McMaster University, Canada
  (2012).

\end{thebibliography}

\end{document}